\documentclass[12pt,a4paper]{article}
\usepackage{amsmath}
\usepackage{amssymb}
\usepackage{tikz}
\usepackage{cite}
\usepackage{hyperref}

\makeatletter
  \def\@seccntformat#1{%
    \@nameuse{@seccnt@prefix@#1}%
    \@nameuse{the#1}%
    \@nameuse{@seccnt@postfix@#1}%
    \@nameuse{@seccnt@afterskip@#1}}
  \def\@seccnt@prefix@section{}
  \def\@seccnt@postfix@section{.}
  \def\@seccnt@afterskip@section{\hspace{.5em}}
  \def\@seccnt@prefix@subsection{}
  \def\@seccnt@postfix@subsection{.}
  \def\@seccnt@afterskip@subsection{\hspace{.5em}}
\makeatother

\makeatletter
\renewcommand\section{
  \@startsection{section}{3}{\z@}%
  {-3.25ex\@plus -1ex \@minus -.2ex}%
  {1.5ex \@plus .2ex}%
  {\normalfont\normalsize\bfseries\mathversion{bold}}}
\renewcommand\subsection{
  \@startsection{subsection}{3}{\z@}%
  {-3.25ex\@plus -1ex \@minus -.2ex}%
  {1.5ex \@plus .2ex}%
  {\normalfont\normalsize\bfseries\mathversion{bold}}}
\makeatother

\makeatletter \@addtoreset{equation}{section} \makeatother
\renewcommand{\theequation}{\arabic{section}.\arabic{equation}}

\allowdisplaybreaks[1]

\let\oldthebibliography\thebibliography
\renewcommand\thebibliography[1]{
  \oldthebibliography{#1}\setlength{\itemsep}{0.4ex}}

\newcommand{\nn}{\nonumber}

\newcommand{\alg}[1]{\mathfrak{#1}}
\newcommand{\grp}[1]{\mathrm{#1}}

\newcommand{\bbC}{\mathbb{C}}
\newcommand{\bbR}{\mathbb{R}}
\newcommand{\bbZ}{\mathbb{Z}}
\newcommand{\bbH}{\mathbb{H}}
\newcommand{\varth}{\vartheta}

\newcommand{\tc}{\tilde{c}}
\newcommand{\tf}{\tilde{f}}
\newcommand{\tg}{\tilde{g}}
\newcommand{\tm}{\widetilde{m}}
\newcommand{\tp}{\tilde{p}}
\newcommand{\tu}{\tilde{u}}
\newcommand{\tz}{\tilde{z}}
\newcommand{\tphi}{\tilde{\phi}}

\newcommand{\cS}{\mathcal{S}}
\newcommand{\cT}{\mathcal{T}}

\newcommand{\vecm}{{\boldsymbol{m}}}
\newcommand{\vecs}{{\boldsymbol{s}}}
\newcommand{\vecv}{{\boldsymbol{v}}}
\newcommand{\vecw}{{\boldsymbol{w}}}
\newcommand{\vecz}{{\boldsymbol{z}}}
\newcommand{\vecal}{{\boldsymbol{\alpha}}}
\newcommand{\vecbe}{{\boldsymbol{\beta}}}
\newcommand{\vece}{{\boldsymbol{\mathrm{e}}}}
\newcommand{\vecmu}{{\boldsymbol{\mu}}}
\newcommand{\vecom}{{\boldsymbol{\omega}}}
\newcommand{\vecL}{{\boldsymbol{\Lambda}}}
\newcommand{\veczero}{{\boldsymbol{0}}}

\newcommand{\Rt}{R^\vee}
\newcommand{\bfR}{\boldsymbol{R}}

\newcommand{\hlf}{\mathrm{h}}
\newcommand{\dbl}{\mathrm{d}}
\newcommand{\tri}{*}

\newcommand{\DGHKKL}{{\mbox{\tiny{DGHKKL}}}}

\newcommand{\Gtri}{{\mathcal{G}}}

\begin{document}


\def\papertitlepage{\baselineskip 3.5ex \thispagestyle{empty}}
\def\preprinumber#1#2{\hfill
\begin{minipage}{1.22in}
#1 \par\noindent #2
\end{minipage}}

%
\papertitlepage
\setcounter{page}{0}
\preprinumber{arXiv:2304.04878}{}
\vskip 2ex
\vfill
\begin{center}
{\large\bf\mathversion{bold}
E-strings, $F_4$, and $D_4$ triality
}
\end{center}
\vfill
\baselineskip=3.5ex
\begin{center}
Kazuhiro Sakai\\

{\small
\vskip 6ex
{\it Institute of Physics, Meiji Gakuin University,
Yokohama 244-8539, Japan}\\
\vskip 1ex
{\tt kzhrsakai@gmail.com}

}
\end{center}
\vfill
\baselineskip=3.5ex
\begin{center} {\bf Abstract} \end{center}

We study the E-string theory on $\mathbb{R}^4\times T^2$
with Wilson lines. We consider two examples where interesting
automorphisms arise. In the first example, the spectrum is invariant
under the $F_4$ Weyl group acting on the Wilson line parameters.
We obtain the Seiberg--Witten curve expressed in terms of
Weyl invariant $F_4$ Jacobi forms. We also clarify how it is
related to the thermodynamic limit of the Nekrasov-type formula.
In the second example, the spectrum is invariant under the $D_4$
triality combined with modular transformations, the automorphism
originally found in the 4d $\mathcal{N}=2$ supersymmetric
$\mathrm{SU}(2)$ gauge theory with four massive flavors.
We introduce the notion of triality invariant Jacobi forms
and present the Seiberg--Witten curve in terms of them.
We show that this Seiberg--Witten curve reduces precisely
to that of the 4d theory with four flavors
in the limit of $T^2$ shrinking to zero size.

\vfill
\noindent
April 2023


\setcounter{page}{0}
\newpage
\renewcommand{\thefootnote}{\arabic{footnote}}
\setcounter{footnote}{0}
\setcounter{section}{0}
\baselineskip = 3.5ex
\pagestyle{plain}

\section{Introduction}\label{sec:intro}

The E-string theory is a fundamental superconformal field theory (SCFT)
in six dimensions \cite{Ganor:1996mu,Seiberg:1996vs}.
It is commonly defined as the low energy theory of
a coincident M5--M9 brane system.
The theory preserves eight supercharges and possesses
a global $E_8$ symmetry.
When toroidally compactified down to four dimensions,
it admits the Seiberg--Witten description
\cite{Seiberg:1994rs,Seiberg:1994aj}.
Upon compactification
one can introduce Wilson lines for the global $E_8$ symmetry.
The Wilson lines are specified by eight parameters
$m_1,\ldots, m_8$
corresponding to the Cartan part of the $E_8$,
which are viewed as the mass parameters of fundamental matters
in the low-energy gauge theory \cite{Ganor:1996xd,Ganor:1996pc}.

By tuning these parameters $m_i$ together with
the modulus $\tau$ and the size $L$ of the torus,
the E-string theory can flow to almost all
rank-one gauge theories and SCFTs
preserving eight supercharges in five and four dimensions
\cite{Ganor:1996xd,Ganor:1996pc,Minahan:1997ch,Eguchi:2002fc,
Eguchi:2002nx}.
Among others, of particular interest is the flow to
the 4d $\mathcal{N}=2$ supersymmetric $\grp{SU(2)}$ gauge theory
with four massive flavors ($N_\mathrm{f}=4$ theory).
This theory is special in many respects \cite{Seiberg:1994aj},
especially it retains the infrared gauge coupling $\tau_\mathrm{IR}$
as a parameter.
A natural identification of the parameters 
in the above flow is that
the modulus $\tau$ of the torus turns into $\tau_\mathrm{IR}$
while four out of eight Wilson line parameters $m_i$
give the masses of 4d flavor matters.

However, there is a puzzle concerning this reduction.
There are known various ways to obtain
the 4d $N_\mathrm{f}=4$ theory from the E-string theory,
but none of them realizes the above parameter identification.
For instance, one can first reduce the E-string theory to
the (mass-deformed) 4d Minahan--Nemeschansky SCFT of $E_8$ type
\cite{Minahan:1996cj}
and then further reduce it to
obtain the $N_\mathrm{f}=4$ theory. In this case,
$m_i$ are trivially identified,
but $\tau$ is sent to a special value
while $\tau_\mathrm{IR}$ is pulled out from the extra
mass parameters of the 4d $E_8$ theory.
In another reduction \cite{Sakai:2012ik},
$\tau$ is identified with $\tau_\mathrm{IR}$ as desired,
but the identification of $m_i$ and the 4d masses is quite non-trivial.

In this paper we study in detail the E-string theory
on $\bbR^4\times T^2$ with four Wilson line parameters
and resolve this puzzle.
One may think that a natural candidate of the Wilson line
configuration that leads to the 4d $N_\mathrm{f}=4$ theory is
\begin{align}
\vecm=\vecm_\hlf:=(m_1,m_2,m_3,m_4,0,0,0,0).
\end{align}
Clearly, this configuration admits the invariance of the spectrum 
under the $D_4$ Weyl group $W(D_4)$ acting on $m_i$,
the symmetry that the massive 4d $N_\mathrm{f}=4$ theory
evidently possesses. As pointed out in \cite{Sakai:2012ik},
this configuration is equivalent to
\begin{align}
\vecm
 =\vecm_\dbl
 :=(\mu_1,\mu_2,\mu_3,\mu_4,\mu_1,\mu_2,\mu_3,\mu_4),
\end{align}
where $m_i$ and $\mu_j$ are related in a simple manner.
The E-string theory with the latter Wilson line configuration has been 
studied, e.g.~in \cite{Sakai:2012ik,Haghighat:2018dwe,Chen:2021ivd}.
Indeed, there is a way to identify the 6d theory with this 
configuration and the 4d $N_\mathrm{f}=4$ theory \cite{Sakai:2012ik},
but this is at the cost of conceding nontrivial
identification of mass parameters, as mentioned above.
What makes the identification so intricate?
The answer we find is that
the symmetry acting non-trivially on the above configuration is
in fact larger than $W(D_4)$:
We point out that the 6d theory admits a $W(F_4)$ automorphism.

Since the 4d $N_\mathrm{f}=4$ theory does not possess
the full $W(F_4)$ symmetry,\footnote{Note, however, that
in the $N_\mathrm{f}=4$ theory physical quantities
on which the modular group trivially acts
can have the $W(F_4)$ invariance: For instance,
the superconformal index is known to be $W(F_4)$ invariant
\cite{vandeBult2009,Gadde:2009kb}. This can also be seen
from the viewpoint of the E-string theory \cite{Kim:2017toz}.}
one has to break it down to $W(D_4)$
at some point of the reduction.
At the same time, it is well known that
the $N_\mathrm{f}=4$ theory has an interesting $D_4$ triality
symmetry: the theory is invariant under the outer automorphism
of the $D_4$ Dynkin diagram combined with modular transformations
\cite{Seiberg:1994aj}.
How and when does this symmetry arise from the 6d theory?
We find that there exists a purely 6d setup
where just enough symmetry is already present before reduction.
This can be achieved by adding a further twist
to the Wilson line parameters.
There are several equivalent variations,
but one of such twisted configurations is
\begin{align}
\vecm
 =\vecm_\tri
 :=(m_1,m_2,m_3,m_4,
 \tfrac{1}{2},-\tfrac{1}{2},-\tfrac{1}{2},-\tfrac{1}{2}-\tau).
\end{align}
We show that the E-string theory with this configuration
admits the above $D_4$ triality symmetry.

Another aim of this paper is to investigate
a fully symmetric description of physical quantities
in the E-string theory with the above configurations.
We concentrate on the Seiberg--Witten curve,
since it determines various important 
supersymmetric indices directly or indirectly.
Apart from technical complication, a fully symmetric description
can be obtained straightforwardly for the $F_4$ configurations
$\vecm=\vecm_\hlf$ and $\vecm=\vecm_\dbl$:
We present the Seiberg--Witten curve expressed in terms of
the $W(F_4)$-invariant Jacobi forms in both cases.

The case of $\vecm=\vecm_\tri$ is more interesting.
As expected, one can write the Seiberg--Witten curve
in terms of the $W(D_4)$-invariant Jacobi forms,
but this is not the end of the story.
As mentioned above, the theory in addition
exhibits the $D_4$ triality symmetry.
To take this symmetry into account,
we introduce the notion of the $D_4$ triality invariant Jacobi forms
and present the Seiberg--Witten curve in terms of them.
As far as we know, Jacobi forms of this kind
have never been considered in the literature before.
Further investigation of them may provide us with new insights into
the study of Jacobi forms.

We also study the reductions of these theories down to
five and four dimensions.
For the $W(F_4)$-invariant theory in five and four dimensions,
we write the Seiberg--Witten curve in terms of $F_4$ Weyl orbit
characters and $F_4$ Casimir invariants respectively.
For the $D_4$ triality invariant theory in the 5d limit,
we find that the Seiberg--Witten curve degenerates 
everywhere in the Coulomb branch moduli space,
meaning that the theory becomes trivial.
In the 4d limit, the theory is expected to flow to
the $N_\mathrm{f}=4$ theory. This is indeed the case.
We explicitly show that the Seiberg--Witten curve
written in terms of triality invariant Jacobi forms
reduces precisely to that of the 4d theory
in the limit of $T^2$ shrinking to zero size.

We also clarify how the Seiberg--Witten curve
for the $W(F_4)$-invariant model is related to
the thermodynamic limit of the Nekrasov-type formula
for the prepotential proposed in \cite{Sakai:2012ik}.
The Nekrasov-type formula is a special case of
the 6d generalization \cite{Hollowood:2003cv}
of the Nekrasov partition function
\cite{Nekrasov:2002qd,Nekrasov:2003rj}, whose 
thermodynamic limit has been well studied \cite{Hollowood:2003cv}.
However, the resulting Seiberg--Witten curve is commonly
expressed in a non-elliptic form.
We describe in detail how to transform it into a quartic elliptic curve
and subsequently into a cubic Weierstrass form,
which precisely reproduces our $F_4$ curve.
This completes the proof of the Nekrasov-type formula
for the E-string theory,
which has been done only in certain special cases \cite{Ishii:2013nba}.

The paper is organized as follows.
In section~\ref{sec:F4} we study the $W(F_4)$-invariant model.
We first show the emergence of the $W(F_4)$ symmetry
and then present the Seiberg--Witten curve in terms of
$W(F_4)$-invariant Jacobi forms.
We also study the 5d and 4d limits.
In section~\ref{sec:Nek}
we study the thermodynamic limit of the Nekrasov-type formula
for the E-string theory
and show its equivalence with the $F_4$ curve.
Along the way, we present two simple expressions
for the $F_4$ curve.
Section~\ref{sec:D4tri} is devoted to
the $D_4$ triality invariant model.
We first show that the E-string theory with the twisted Wilson line
configuration admits the $D_4$ triality symmetry.
We then clarify the automorphism group that includes the triality
symmetry and introduce the notion of triality invariant Jacobi forms.
We present the Seiberg--Witten curve written in terms of them
and consider its 5d and 4d limits. In particular, we explicitly show
that it reproduces the known Seiberg--Witten curve for
the $N_\mathrm{f}=4$ theory.
In section~\ref{sec:conclusion}
we summarize our results and discuss possible directions
for further studies.
The definitions of special functions and
some useful formulas are presented in Appendix~\ref{app:functions}.

\section{$W(F_4)$-invariant model}\label{sec:F4}

\subsection{Emergence of $W(F_4)$ automorphism}\label{sec:F4auto}

The E-string theory is a 6d SCFT and has a global $E_8$ symmetry.
When the theory is partially compactified on $T^2$,
one can introduce Wilson lines for the $E_8$ symmetry.
The Wilson lines are specified
by eight Wilson line parameters 
\begin{align}
\vecm=(m_1,\ldots,m_8),\qquad m_i=p_i+q_i\tau\in\bbC.
\end{align}
Here $p_i$ and $q_i$ are $E_8$ Cartan charges
along the fifth and sixth directions respectively
and $\tau$ is the complex structure modulus of the torus.
We fix the size of the torus in such a way that
$\vecm$ satisfies the periodicity condition described below.
Thus the E-string theory on $\bbR^4\times T^2$
with Wilson lines is specified by
the coordinate
\begin{align}
(\tau,\vecm)
\end{align}
in the parameter space $\bbH\times\bbC^8$,
where $\bbH=\{\tau\in\bbC|\mathrm{Im}\,\tau>0\}$
is the upper half plane.

However, the above parametrization is of course redundant and
different points on the parameter space are in fact
identified with one another.
In other words, 
the parameter space has non-trivial automorphisms.
For general $\vecm$
the theory exhibits three kinds of automorphisms \cite{Ganor:1996xd}:
\renewcommand{\theenumi}{\roman{enumi}}
\renewcommand{\labelenumi}{(\theenumi)}
\begin{enumerate}
\item $E_8$ Weyl group automorphism:
\begin{align}
(\tau,w(\vecm)) \sim (\tau,\vecm),\qquad
w\in W(E_8).
\label{eq:WeylE8inv}
\end{align}

\item Double periodicity:
\begin{align}
(\tau,\vecm+\tau\vecal+\vecbe)
\sim (\tau,\vecm),\qquad
\vecal,\vecbe\in L_{E_8}.
\label{eq:periodicity}
\end{align}

\item $\grp{SL(2,\bbZ)}$ automorphism:
\begin{align}
\left(
\frac{a\tau+b}{c\tau+d}\,,\frac{\vecm}{c\tau+d}\right)
\sim (\tau,\vecm),\qquad
\left(\begin{array}{cc}a&b\\ c&d\end{array}\right)
\in \grp{SL}(2,\bbZ).
\label{eq:SL2Zinv}
\end{align}
\end{enumerate}
Here $W(E_8)$ and $L_{E_8}$ denote the Weyl group
and the root lattice of $E_8$ respectively.
In the rest of this subsection let us look into
the Weyl group automorphism in more detail.

Let $\vece_i\ (i=1,\ldots,8)$ be the standard basis of $\bbR^8$.
The roots of $E_8$ are given by
\begin{align}
\begin{aligned}
&\pm\vece_i\pm\vece_j\quad(i<j),\\
&\frac{1}{2}\left(\pm\vece_1\cdots\pm\vece_8\right)
\quad\mbox{with even number of `+'}.
\end{aligned}
\end{align}
The Weyl group of $E_8$ is generated by the reflections
with respect to these roots.
Let $s_\vecal$ denote the reflection with respect to
the root $\vecal$. Specifically, $s_\vecal$ acts on $\vecm$ as
\begin{align}
s_\vecal(\vecm)=\vecm-2\frac{\vecal\cdot\vecm}{\vecal^2}\vecal.
\end{align}
For instance, using the expression $\vecm=\sum_{i=1}^8 m_i\vece_i$
we have
\begin{align}
\begin{aligned}
s_{\vece_1-\vece_2}(\vecm)
 &=\vecm-(m_1-m_2)(\vece_1-\vece_2)\\
 &=m_2\vece_1+m_1\vece_2+\sum_{i=3}^8 m_i\vece_i.
\end{aligned}
\end{align}
From this we see
that $s_{\vece_i-\vece_j}$ interchanges the $i$th and $j$th
components of $\vecm$.
Let us express this reflection as
\begin{align}
s_{\vece_i-\vece_j}:\quad m_i\leftrightarrow m_j.
\label{eq:WeylRef1}
\end{align}
Similarly, we have
\begin{align}
s_{\vece_i+\vece_j}:\quad m_i\leftrightarrow -m_j.
\label{eq:WeylRef2}
\end{align}
Note that the combination of the above two reflections yields
the simultaneous change of two signs
\begin{align}
s_{\vece_i+\vece_j}s_{\vece_i-\vece_j}:\quad
m_i\to -m_i,\quad m_j\to -m_j.
\label{eq:WeylRef12}
\end{align}
In a general context,
reflections \eqref{eq:WeylRef1} and \eqref{eq:WeylRef2}
correspond to the $D_n$ roots
$\vece_i\pm\vece_j\ (1\le i\ne j\le n)$
and thus generate the Weyl group $W(D_n)$.
In the present case
they generate $W(D_8)$. In addition, for the root vector
\begin{align}
\vecs=\frac{1}{2}\sum_{j=1}^8\vece_j,
\end{align}
we have
\begin{align}
s_\vecs:\quad m_i\to m_i-\frac{1}{4}\sum_{j=1}^8m_j\quad
  \mbox{for all $i$}.
\label{eq:WeylRef3}
\end{align}
It is well known that the reflections
\eqref{eq:WeylRef1}, \eqref{eq:WeylRef2} and \eqref{eq:WeylRef3}
generate the full Weyl group $W(E_8)$.

Let us now consider the situation where the Wilson line parameters
are partially
fixed as
\begin{align}
\vecm=\vecm_\hlf:=(m_1,m_2,m_3,m_4,0,0,0,0).
\label{eq:F4config1}
\end{align}
As long as this constraint is imposed,
the parameter space is effectively $\bbH\times\bbC^4$.
What is the nontrivial Weyl group automorphism acting on this space?
From the above discussion it is obvious that
there is a Weyl group of $D_4$ acting on $m_1,\ldots,m_4$,
which is a subgroup of $W(E_8)$.
However, as announced repeatedly,
we have in fact a slightly larger automorphism.
In what follows we will show that
the parameter space $\bbH\times\bbC^4$ admits
a $W(F_4)$ automorphism.

The roots of $F_4$ are given by
\begin{align}
\begin{aligned}
\mbox{long roots:}\quad&\pm\vece_i\pm\vece_j\quad (1\le i<j\le 4),\\
\mbox{short roots:}\quad&\pm\vece_i\quad (i=1,2,3,4),\qquad
\frac{1}{2}\left(\pm\vece_1\pm\vece_2\pm\vece_3\pm\vece_4\right).
\end{aligned}
\label{eq:F4roots1}
\end{align}
The long roots are identical with the $D_4$ roots,
from which it is obvious that $W(D_4)\subset W(F_4)$.
With respect to short roots,
$\vecm_\hlf$ is reflected as, for instance,
\begin{align}
\begin{aligned}
s_{\vece_4}(\vecm_\hlf)&=(m_1,m_2,m_3,-m_4,0,0,0,0),\\
s_{\frac{1}{2}(\vece_1+\vece_2+\vece_3+\vece_4)}(\vecm_\hlf)
&=\vecm_\hlf',
\end{aligned}
\label{eq:F4refl}
\end{align}
where
\begin{align}
\vecm_\hlf'=(
 \tfrac{m_1-m_2-m_3-m_4}{2},
 \tfrac{-m_1+m_2-m_3-m_4}{2},
 \tfrac{-m_1-m_2+m_3-m_4}{2},
 \tfrac{-m_1-m_2-m_3+m_4}{2},
 0,0,0,0).
\end{align}
However, these results are also obtained by $W(E_8)$ reflections as
\begin{align}
\begin{aligned}
(s_{\vece_4-\vece_5}s_{\vece_4+\vece_5})(\vecm_\hlf)
 &=(m_1,m_2,m_3,-m_4,0,0,0,0),\\
(s_\vecs
s_{\vece_7-\vece_8}
s_{\vece_7+\vece_8}
s_{\vece_5-\vece_6}
s_{\vece_5+\vece_6}
s_\vecs)(\vecm_\hlf)
 &=\vecm_\hlf'.
\end{aligned}
\label{eq:virtualF4}
\end{align}
That is, these $W(E_8)$ reflections
virtually act as $W(F_4)$ reflections on $\vecm_\hlf$.
More generally, one can choose the simple roots of $F_4$ as
\begin{align}
\vece_2-\vece_3,\quad
\vece_3-\vece_4,\quad
\vece_4,\quad
-\frac{1}{2}(\vece_1+\vece_2+\vece_3+\vece_4).
\end{align}
Therefore, the reflections \eqref{eq:virtualF4}, together with
$s_{\vece_2-\vece_3}$ and $s_{\vece_3-\vece_4}$,
generate the full Weyl group $W(F_4)$.

The above discussion can be applied
to any space on which $W(E_8)$ automorphism group acts,
such as the moduli spaces of field theories
with $E_8$ global/gauge symmetries.
In this paper we will study the examples of
the E-string theory and its reductions to five and four dimensions.

\subsection{Weyl invariant Jacobi forms}\label{sec:jacobi}

Jacobi forms are functions which have characteristics
of both elliptic functions and modular forms.
In particular, Weyl invariant Jacobi forms provide us with
a natural language 
in describing physical quantities
when the system possesses the three kinds of symmetries
\eqref{eq:WeylE8inv}--\eqref{eq:SL2Zinv}.
In this subsection we recall
the definition of the Weyl invariant Jacobi forms
and present the concrete generators that we will use.

Let $R$ be an irreducible root system of rank $r$ and $W(R)$ the Weyl
group of $R$. Let $L_R$ be the root lattice of $R$ and $L_R^*$ the
dual lattice of $L_R$. When $L_R$ is odd, we rescale its bilinear form
by 2 so that it becomes an even lattice. A holomorphic function 
$\varphi_{k,m}(\tau,\vecz)\ (\tau\in\bbH,\ \vecz\in\bbC^r)$
is called a $W(R)$-invariant weak Jacobi form of weight $k$ and
index $m$ ($k\in\bbZ,\ m\in\bbZ_{\ge 0}$) if it satisfies the
following properties \cite{EichlerZagier, Wirthmuller}:
\renewcommand{\theenumi}{\roman{enumi}}
\renewcommand{\labelenumi}{(\theenumi)}
\begin{enumerate}
\item Weyl invariance:
\begin{align}
\varphi_{k,m}(\tau,w(\vecz))
 =\varphi_{k,m}(\tau,\vecz),\qquad w\in W(R).
\label{Weylinv}
\end{align}

\item Quasi-periodicity:
\begin{align}
\varphi_{k,m}(\tau,\vecz+\tau\vecal+\vecbe)
 =e^{-m \pi i (\tau\vecal^2+2\vecz\cdot\vecal)}
  \varphi_{k,m}(\tau,\vecz),\qquad \vecal,\vecbe\in L_R.
\label{quasiperiod}
\end{align}

\item Modular transformation law:
\begin{align}\label{Modularprop}
&
\varphi_{k,m}\left(
  \frac{a\tau+b}{c\tau+d}\,,\frac{\vecz}{c\tau+d}\right)
 =(c\tau+d)^k\exp\left(m\pi i\frac{c}{c\tau+d}\,\vecz^2\right)
  \varphi_{k,m}(\tau,\vecz),\\[1ex]
&
\Bigl(\begin{array}{cc}a&b\\ c&d\end{array}\Bigr)
 \in\grp{SL}(2,\bbZ).\nn
\end{align}

\item $\varphi_{k,m}(\tau,\vecz)$ admits
a Fourier expansion of the form
\begin{align}
\varphi_{k,m}(\tau,\vecz)
=\sum_{n=0}^\infty\sum_{\vecw\in L_R^*}
  c(n,\vecw)e^{2\pi i\vecw\cdot\vecz}q^n.
\label{Fourierform}
\end{align}
\end{enumerate}
If $\varphi_{k,m}(\tau,\vecz)$ further satisfies the condition
that the coefficients $c(n,\vecw)$ of the Fourier
expansion (\ref{Fourierform}) vanish unless $\vecw^2\le 2mn$, it is
called a $W(R)$-invariant holomorphic Jacobi form. In this paper
a Jacobi form means a weak Jacobi form unless otherwise specified.

Let
\begin{align}
J^{R}_{k,m}
\end{align}
denote the vector space of $W(R)$-invariant Jacobi forms of
weight $k$ and index $m$.
The space of all $W(R)$-invariant Jacobi forms
\begin{align}
J^{R}_{*,*}
 :=\bigoplus_{k\in\bbZ,\,m\in\bbZ_{\ge 0}}J^{R}_{k,m}
\end{align}
is a bigraded algebra over the ring of modular forms.
It is known that for any irreducible root system $R$
not of type $E_8$, $J^{R}_{*,*}$ is a polynomial algebra
generated by $r+1$ generators \cite{Wirthmuller}.
This means that every $W(R)$-invariant Jacobi form
is expressed uniquely as a polynomial of
$r+1$ generators and the Eisenstein series
$E_4(\tau),E_6(\tau)$.
The generators have been explicitly constructed
for all types (except $E_8$) of root systems
\cite{Wirthmuller,Satake:1993cp,BertolaThesis,Bertola1,
Sakai:2017ihc,Adler:2019ysr,Adler:2021}.
On the other hand, the ring $J^{E_8}_{*,*}$ is not a polynomial
algebra \cite{Wang:2018fil} and has a complicated structure
\cite{Sun:2021ije,Sakai:2022taq}.
Nevertheless, one can systematically construct
these Jacobi forms using nine ``generators''
$A_i,B_j$ or $a_i,b_j$ \cite{Sun:2021ije,Sakai:2022taq}.
In the rest of this subsection
we explicitly present the generators
for $R=D_4,F_4,E_8$, which we will use later.

\paragraph{\mathversion{bold}$D_4$}

The generators of the ring of $W(D_4)$-invariant Jacobi forms
were constructed in \cite{Wirthmuller,BertolaThesis}.
In particular, fully explicit forms were
given by Bertola \cite{BertolaThesis}.
In this paper we adopt the notation of
the reference \cite{DelZotto:2017mee},
where Bertola's generators
are expressed more explicitly
(with slight change of normalization)
in terms of well-known functions.
The five generators are of the following weights and indices
\begin{align}
\varphi_0\in J^{D_4}_{0,1},\quad
\varphi_2\in J^{D_4}_{-2,1},\quad
\varphi_4\in J^{D_4}_{-4,1},\quad
\psi_4   \in J^{D_4}_{-4,1},\quad
\varphi_6\in J^{D_4}_{-6,2}
\end{align}
and are given by
\begin{align}
\begin{aligned}
\varphi_0
 &=\eta^{-12}\left[
   \varth_3^8T_3-\varth_4^8T_4-\varth_2^8T_2\right],\\
\varphi_2
 &=\eta^{-12}\left[
   (\varth_4^4-\varth_2^4)T_3
  -(\varth_2^4+\varth_3^4)T_4
  +(\varth_3^4+\varth_4^4)T_2
 \right],\\
\varphi_4
 &=\eta^{-12}\left[T_3-T_4-T_2\right],\\
\psi_4
 &=\eta^{-12}T_1,\\
\varphi_6
 &=\pi^{-2}\eta^{-24}T_1^2
 \textstyle\sum_{j=1}^{4}\wp(m_j)
\end{aligned}
\label{eq:D4gen}
\end{align}
with\footnote{By abuse of notation we let $\vecm$ denote
8-vector for $E_8$ and 4-vector for $D_4$ and $F_4$.}
\begin{align}
T_k(\tau,\vecm) := \prod_{j=1}^4\varth_k(m_j,\tau).
\label{eq:Tdef}
\end{align}
Here $\eta=\eta(\tau)$, $\varth_k(z,\tau)$ and $\wp(z)$ are
the Dedekind eta function,
the Jacobi theta functions and the Weierstrass elliptic function
respectively (see Appendix~\ref{app:functions} for our convention).
In this paper we often abbreviate $\varth_k(0,\tau)$ as $\varth_k$.

\paragraph{\mathversion{bold}$F_4$}

The generators of the ring of $W(F_4)$-invariant Jacobi forms
were constructed in \cite{Wirthmuller,Adler:2021}.
In particular, fully explicit forms were
given by Adler \cite{Adler:2021}.
Here we follow the convention of \cite{Duan:2020imo}.
The five generators are of the following weights and indices
\begin{align}
\phi_0\in J^{F_4}_{0,1},\quad
\phi_2\in J^{F_4}_{-2,1},\quad
\phi_6\in J^{F_4}_{-6,2},\quad
\phi_8\in J^{F_4}_{-8,2},\quad
\phi_{12}\in J^{F_4}_{-12,3}
\end{align}
and are expressed in terms of
the $D_4$ generators \eqref{eq:D4gen} as\footnote{There is
a small difference between the definition of the $D_4$ generators 
in \cite{DelZotto:2017mee}, which we use in this paper,
and that in \cite{Duan:2020imo}:
generators in the former and latter are related as
\begin{align*}
\varphi_0=\phi^{D_4}_{0,1},\quad
\varphi_2=\phi^{D_4}_{-2,1},\quad
\varphi_4=\phi^{D_4}_{-4,1},\quad
\psi_4=\omega^{D_4}_{-4,1},\quad
\varphi_6=4\phi^{D_4}_{-6,2}.
\end{align*}
}
\begin{align}
\begin{aligned}
\phi_0
 &=\varphi_0-\frac{2}{3}E_4\varphi_4,\\
\phi_2
 &=\varphi_2,\\
\phi_6
 &=\frac{1}{4}\varphi_6-\frac{1}{18}\varphi_2\varphi_4,\\
\phi_8
 &=\varphi_4^2+3\psi_4^2,\\
\phi_{12}
 &=\varphi_4\psi_4^2-\frac{1}{9}\varphi_4^3.
\end{aligned}
\label{eq:F4gen}
\end{align}

We remind the reader that these $W(F_4)$-invariant Jacobi forms are 
constructed for the rescaled root lattice
\begin{align}
L_{F_4}=L_{D_4}
 &:=\bigl\{\vecom=(\omega_1,\omega_2,\omega_3,\omega_4)\in\bbZ^4
  \bigm| \textstyle\sum_{j=1}^4 \omega_j\in 2\bbZ\bigr\}.
\label{eq:rescaledLF4}
\end{align}
Correspondingly, the rescaled $F_4$ roots are given by
\begin{align}
\begin{aligned}
\mbox{long roots:}\quad&\pm 2\vece_i\quad (i=1,2,3,4),\qquad
\pm\vece_1\pm\vece_2\pm\vece_3\pm\vece_4,\\
\mbox{short roots:}\quad&\pm\vece_i\pm\vece_j\quad (1\le i<j\le 4),
\end{aligned}
\label{eq:F4roots2}
\end{align}
rather than \eqref{eq:F4roots1}.
Note that this change of the basis does not alter
how the whole $W(F_4)$ acts on $(m_1,\ldots,m_4)$,
nor does it harm the discussion in the last subsection.

\paragraph{\mathversion{bold}$E_8$}

The study of $W(E_8)$-invariant Jacobi forms was initiated
in \cite{Minahan:1998vr}.
Although $W(E_8)$-invariant Jacobi forms do not form
a polynomial algebra over the ring of modular forms,
they are still expressed as polynomials of nine independent generators
if one allows modular functions, rather than modular forms,
as coefficients \cite{Sakai:2017ihc,Wang:2018fil}
(see \cite{Sun:2021ije,Sakai:2022taq} for systematic constructions
of such polynomials).
A natural choice of the nine generators
is to take $A_n\ (n=1,2,3,4,5)$, $B_n\ (n=2,3,4,6)$
constructed in \cite{Sakai:2011xg}.
They are $W(E_8)$-invariant holomorphic Jacobi forms
of the following weights and indices
\begin{align}
A_n\in J^{E_8}_{4,n},\quad
B_n\in J^{E_8}_{6,n}
\end{align}
and are given by
\begin{align}
A_1(\tau,\vecm)&=\Theta_{E_8}(\tau,\vecm),\qquad
A_4(\tau,\vecm)=A_1(\tau,2\vecm),\nn\\
A_n(\tau,\vecm)&=\tfrac{n^3}{n^3+1}\left(
 A_1(n\tau,n\vecm)
  +\tfrac{1}{n^4}\mbox{$\sum_{k=0}^{n-1}$}A_1(\tfrac{\tau+k}{n},\vecm)
  \right),\qquad n=2,3,5,\nn\\
B_2(\tau,\vecm)&=\tfrac{32}{5}\left(
 e_1(\tau)A_1(2\tau,2\vecm)
  +\tfrac{1}{2^4}e_3(\tau)A_1(\tfrac{\tau}{2},\vecm)
  +\tfrac{1}{2^4}e_2(\tau)A_1(\tfrac{\tau+1}{2},\vecm)\right),\nn\\
B_3(\tau,\vecm)&=\tfrac{81}{80}\left(
 h_0(\tau)^2A_1(3\tau,3\vecm)
  -\tfrac{1}{3^5}\mbox{$\sum_{k=0}^{2}$}h_0(\tfrac{\tau+k}{3})^2
  A_1(\tfrac{\tau+k}{3},\vecm)\right),\nn\\
B_4(\tau,\vecm)&=\tfrac{16}{15}\left(
 \varth_4(2\tau)^4A_1(4\tau,4\vecm)
  -\tfrac{1}{2^4}\varth_4(2\tau)^4
  A_1(\tau+\tfrac{1}{2},2\vecm)\right.\nn\\
&\hspace{2em}
 \left.
 -\tfrac{1}{2^2\cdot 4^4}\mbox{$\sum_{k=0}^{3}$}
  \varth_2(\tfrac{\tau+k}{2})^4
  A_1(\tfrac{\tau+k}{4},\vecm)\right),\nn\\
B_6(\tau,\vecm)&=\tfrac{9}{10}\left(
  h_0(\tau)^2A_1(6\tau,6\vecm)
 +\tfrac{1}{2^4}\mbox{$\sum_{k=0}^{1}$}
  h_0(\tau+k)^2A_1(\tfrac{3\tau+3k}{2},3\vecm)\right.\nn\\
&\hspace{2em}\left.
 -\tfrac{1}{3\cdot 3^4}\mbox{$\sum_{k=0}^{2}$}
  h_0(\tfrac{\tau+k}{3})^2A_1(\tfrac{2\tau+2k}{3},2\vecm)\right.\nn\\
&\hspace{2em}\left.
 -\tfrac{1}{3\cdot 6^4}\mbox{$\sum_{k=0}^{5}$}
  h_0(\tfrac{\tau+k}{3})^2A_1(\tfrac{\tau+k}{6},\vecm)\right).
\label{E8AB}
\end{align}
Here, $\Theta_{E_8}$ is the theta function of the $E_8$ root lattice
\begin{align}
\begin{aligned}
\Theta_{E_8}(\tau,\vecm)
 :=\sum_{\vecw\in L_{E_8}}
   \exp\left(\pi i\tau\vecw^2
   +2\pi i\vecm\cdot\vecw\right)
 =\frac{1}{2}\sum_{k=1}^4\prod_{j=1}^8\varth_k(m_j,\tau),
\end{aligned}
\end{align}
$\varth_k(\tau)=\varth_k(0,\tau)$ and
the functions $e_j(\tau),\ h_0(\tau)$ are defined as
\begin{align}
\begin{aligned}
e_1(\tau)&:=
 \tfrac{1}{12}\left(\varth_3(\tau)^4+\varth_4(\tau)^4\right),\\
e_2(\tau)&:=
 \tfrac{1}{12}\left(\varth_2(\tau)^4-\varth_4(\tau)^4\right),\\
e_3(\tau)&:=
 \tfrac{1}{12}\left(-\varth_2(\tau)^4-\varth_3(\tau)^4\right),\\
h_0(\tau)&:=
 \varth_3(2\tau)\varth_3(6\tau)+\varth_2(2\tau)\varth_2(6\tau).
\end{aligned}
\label{eq:ehdef}
\end{align}
The normalizations of $A_n,B_n$ are determined so that
\begin{align}
A_n(\tau,\veczero)=E_4(\tau),\qquad
B_n(\tau,\veczero)=E_6(\tau).
\end{align}
%

\subsection{$W(F_4)$-invariant Seiberg--Witten curve}

In this subsection we determine the Seiberg--Witten curve
for the E-string theory with 
the partially fixed Wilson line configuration $\vecm=\vecm_\hlf$
given in \eqref{eq:F4config1}.
As we saw in section~\eqref{sec:F4auto},
the theory in this case admits the $W(F_4)$ automorphism.
We call this Seiberg--Witten curve
the $F_4$ curve hereafter.

We start with the manifestly $W(E_8)$-invariant
Seiberg--Witten curve for general $\vecm$
constructed in \cite{Eguchi:2002fc}.
It takes the form
\begin{align}
\begin{aligned}
y^2=4x^3
 &-\left(\frac{E_4}{12}u^4+a_2u^2+a_3u+a_4\right)x\\
 &-\left(\frac{E_6}{216}u^6+b_1u^5+b_2u^4+b_3u^3+b_4u^2+b_5u+b_6\right),
\end{aligned}
\label{eq:E8curve}
\end{align}
where $a_i=a_i(\tau,\vecm),b_j=b_j(\tau,\vecm)$
are certain meromorphic $W(E_8)$-invariant Jacobi forms.
In \cite{Sakai:2011xg}, $a_i,b_j$ were expressed in terms of
$A_k,B_l$ given in \eqref{E8AB}. For instance,
\begin{align}
\begin{aligned}
b_1&=-\frac{4A_1}{E_4},\qquad
a_2=\frac{6(-E_4A_2+A_1^2)}{E_4\Delta},\qquad
b_2=\frac{5(E_4^2B_2-E_6A_1^2)}{6E_4^2\Delta},\\
a_3&=\frac{-7E_4^2E_6A_3-20E_4^3B_3
  -9E_4E_6A_1A_2+30E_4^2A_1B_2+6E_6A_1^3}{9E_4^2\Delta^2},\\
b_3&=\frac{-7E_4^5A_3-20E_4^3E_6B_3
  -9E_4^4A_1A_2+30E_4^2E_6A_1B_2+(16E_4^3-10E_6^2)A_1^3}
 {108E_4^3\Delta^2}.
\end{aligned}
\label{eq:abinAB}
\end{align}
Here $\Delta=\eta^{24}=(E_4^3-E_6^2)/1728$.
It is a straightforward (though tedious) task to express
$A_i,B_j$ at the value \eqref{eq:F4config1}
in terms of the $W(F_4)$-invariant Jacobi forms
$\phi_k$ given in \eqref{eq:F4gen}.
The results for $A_i,B_i$ with $i=1,2,3$ are as follows:
\begin{align}
A_1&=
 \frac{E_4\phi_0}{48}-\frac{E_6\phi_2}{144},\nn\\
A_2&=
  \frac{E_4\phi_0^2}{2304}
 -\frac{E_6\phi_0\phi_2}{3456}
 +\frac{E_4^2\phi_2^2}{20736}
 -\frac{\Delta\phi_8}{18},\nn\\
B_2&=
  \frac{E_6\phi_0^2}{2304}
 -\frac{E_4^2\phi_0\phi_2}{3456}
 +\frac{E_4E_6\phi_2^2}{20736}
 -\frac{6\Delta\phi_6}{5},\nn\\
A_3&=
  \frac{E_4\phi_0(3\phi_0^2+E_4\phi_2^2)}{331776}
 -\frac{E_6\phi_2(27\phi_0^2+E_4\phi_2^2)}{2985984}
 +\frac{\Delta(\phi_0\phi_8+24\phi_2\phi_6-2E_4\phi_{12})}{672},\nn\\
B_3&=
  \frac{E_6\phi_0(3\phi_0^2+E_4\phi_2^2)}{331776}
 -\frac{E_4^2\phi_0^2\phi_2}{110592}
 +\frac{\Delta(\phi_2^3-216\phi_0\phi_6-E_4\phi_2\phi_8+6E_6\phi_{12})}
       {5760}\nn\\
&\hspace{1em}
 -\frac{E_6^2\phi_2^3}{2985984}.
\label{eq:ABinF4}
\end{align}
Substituting relations like these into the curve \eqref{eq:E8curve}
with \eqref{eq:abinAB}, one obtains the $F_4$ curve.

In fact, there is a better way to obtain the $F_4$ curve,
which does not require
the full data of the form \eqref{eq:ABinF4} for $A_4,B_4,A_5,B_6$:
It is clear that the Wilson line configuration \eqref{eq:F4config1}
only partially breaks the global $E_8$ symmetry
and leaves a $D_4$ symmetry.
This means that the elliptic fibration described by
the Seiberg--Witten curve must have a singular fiber
of Kodaira type $\rm{I}^*_0$.
In other words, the $F_4$ curve must take the form
\begin{align}
y^2=4x^3
 -\left(\frac{E_4}{12}u^2+f_1u+f_2\right)(u-u_0)^2x
 -\left(\frac{E_6}{216}u^3+g_1u^2+g_2u+g_3\right)(u-u_0)^3.
\label{eq:F4canz}
\end{align}
Comparing the $x^1$-part of \eqref{eq:F4canz} with
that of \eqref{eq:E8curve}, one obtains the relations
\begin{align}
\begin{aligned}
f_1&=\frac{E_4u_0}{6},\qquad
f_2=\frac{E_4u_0^2}{4}+a_2,\\
0&=E_4 u_0^3+6a_2u_0+3a_3,\\
0&=2a_2u_0^2+3a_3u_0+4a_4.
\end{aligned}
\label{eq:F4feqs}
\end{align}
By using \eqref{eq:abinAB}--\eqref{eq:ABinF4}
the third equation is rewritten as
\begin{align}
\begin{aligned}
0&=E_4 u_0^3+6a_2u_0+3a_3\\
 &=E_4^{-2}(E_4u_0+\phi_2)
           (E_4^2u_0^2-E_4\phi_2u_0+2E_4\phi_8-2\phi_2^2).
\end{aligned}
\label{eq:u0eq}
\end{align}
From this one obtains
\begin{align}
u_0=-\frac{\phi_2}{E_4}.
\label{eq:u0sol}
\end{align}
In principle, \eqref{eq:u0eq} is a cubic equation in $u_0$
and there are two other solutions; but one can check that
they do not satisfy
the fourth equation in \eqref{eq:F4feqs} and are therefore
ruled out.

Given the explicit form \eqref{eq:u0sol} of $u_0$,
the data \eqref{eq:ABinF4} of $A_i,B_i$ for $i\le 3$
suffice to determine the entire form of the $F_4$ curve.
It turns out that the final result can be expressed
in the following surprisingly simple form:
\begin{align}
y^2=4x^3
 -\left(
    \frac{E_4}{12}\tu^2
   -\frac{\phi_2}{3}\tu
   +\frac{\phi_8}{3}
  \right)\tu^2 x
 -\left(
    \frac{E_6}{216}\tu^3
   -\frac{\phi_0}{12}\tu^2
   -\phi_6\tu
   +\frac{\phi_{12}}{3}
  \right)\tu^3
\label{eq:F4curve}
\end{align}
with
\begin{align}
\tu=u+\frac{\phi_2}{E_4}.
\end{align}

Comparing the curves
\eqref{eq:E8curve} and \eqref{eq:F4curve},
one immediately obtains the expressions of $a_i,b_j$
at the value \eqref{eq:F4config1} in terms of $\phi_k$.
Note that the expressions of $A_i,B_j$
in terms of $a_k,b_l$ are explicitly given in \cite{Sakai:2022taq}.
Using this one can easily obtain
the data of the form \eqref{eq:ABinF4} for $A_4,B_4,A_5,B_6$ as well.

\subsection{Another $W(F_4)$ automorphism}

Let us next consider the configuration
\begin{align}
\vecm
 =\vecm_\dbl
 :=(\mu_1,\mu_2,\mu_3,\mu_4,\mu_1,\mu_2,\mu_3,\mu_4).
\label{eq:F4config2}
\end{align}
Seiberg--Witten curve with this configuration
has been studied \cite{Sakai:2012ik,Haghighat:2018dwe,Chen:2021ivd}.
As pointed out in \cite{Sakai:2012ik}, this configuration is
in fact equivalent to $\vecm_\hlf$ in \eqref{eq:F4config1}:
More specifically,
by the following $W(E_8)$ action \eqref{eq:F4config2} is mapped to
\begin{align}
\begin{aligned}
&(s_{\vece_3+\vece_4}s_\vecs s_{\vece_1-\vece_2}s_{\vece_1+\vece_2}
  s_{\vece_5-\vece_6}s_{\vece_5+\vece_6}s_\vecs s_{\vece_3-\vece_8}
  s_{\vece_1-\vece_6}s_{\vece_4-\vece_6}s_{\vece_4+\vece_6})
 (\vecm_\dbl)\\
&=(\mu_1+\mu_2,\mu_1-\mu_2,\mu_3+\mu_4,\mu_3-\mu_4,
   0,0,0,0).
\end{aligned}
\label{eq:vecmu2m}
\end{align}
Therefore $\vecm_\hlf$ and $\vecm_\dbl$
specify the same physical system under the identification
\begin{align}
m_1=\mu_1+\mu_2,\qquad
m_2=\mu_1-\mu_2,\qquad
m_3=\mu_3+\mu_4,\qquad
m_4=\mu_3-\mu_4.
\label{eq:F4map}
\end{align}

In \cite{Sakai:2012ik} it was clarified
how the Seiberg--Witten curve at the value $\vecm_\dbl$
is related to the original curve
of Seiberg and Witten for the 4d $N_\mathrm{f}=4$ theory
\cite{Seiberg:1994aj} (the curve will also be given
in \eqref{eq:Nf4curve1}--\eqref{eq:Nf4curve2}).
Here we present a different point of view.
It was already pointed out in \cite{Sakai:2012ik} that
the fundamental $W(E_8)$-actions
\eqref{eq:WeylRef1}, \eqref{eq:WeylRef12} and \eqref{eq:WeylRef3},
which generate the full $W(E_8)$,
are expressed for $\vecm=\vecm_\dbl$ as
\begin{align}
\begin{aligned}
&\bullet\ \mu_i\leftrightarrow \mu_j\qquad\mbox{for any}\ i\ne j,\\
&\bullet\ \mu_i\leftrightarrow -\mu_i \qquad\mbox{for any}\ i,\\
&\bullet\ \mu_i\rightarrow \mu_i-\frac{1}{2}\sum_{j=1}^4 \mu_j
  \qquad\mbox{for all}\ i=1,\ldots,4.
\end{aligned}
\label{eq:WeylF4bis}
\end{align}
As we learned in section~\ref{sec:F4auto},
these are the very actions that generate
$W(F_4)$ acting on $(\mu_1,\mu_2,\mu_3,\mu_4)$!
In other words, the system exhibits
two different $W(F_4)$ automorphisms,
one acts on $m_j$ and the other acts on $\mu_j$.
Consequently, the $F_4$ curve \eqref{eq:F4curve}
written in terms of $\phi_k=\phi_k(\tau,m_j)$
can also be expressed in terms of
\begin{align}
\tphi_k=\phi_k(\tau,\mu_j).
\end{align}
The result is as follows:
\begin{align}
y^2=4x^3
 -\left(\frac{E_4}{12}\tu^2+\tf_1\tu+\tf_2\right)\tu^2 x
 -\left(\frac{E_6}{216}\tu^3+\tg_1\tu^2
       +\tg_2\tu+\tg_3\right)\tu^3
\label{eq:F4mucurve}
\end{align}
with
\begin{align}
\tu&=u-\frac{3\tf_1}{E_4},\nn\\
\tf_1
 &=-\frac{1}{216}
  \left(3\tphi_0\tphi_2+36E_4\tphi_6+E_6\tphi_8\right),\nn\\
\tf_2
 &=\frac{1}{15552}
  \left(
   \tphi_2^4
  +36\tphi_0^2\tphi_8
  +216\tphi_0\tphi_2\tphi_6
  +E_4\bigl(-2\tphi_2^2\tphi_8+108\tphi_0\tphi_{12}+1296\tphi_6^2\bigr)
\right.\nn\\
 &\hspace{2em}\left.
  +E_6\bigl(72\tphi_6\tphi_8-36\tphi_2\tphi_{12}\bigr)
  +E_4^2\tphi_8^2
\right),\nn\\
\tg_1
 &=-\frac{1}{5184}
  \left(9\tphi_0^2+E_4\tphi_2^2+72E_6\tphi_6+2E_4^2\tphi_8\right),\nn\\
\tg_2
 &=\frac{1}{93312}
  \left(
   3\tphi_0\tphi_2^3+324\tphi_0^2\tphi_6
  +E_4\bigl(15\tphi_0\tphi_2\tphi_8+36\tphi_2^2\tphi_6\bigr)
\right.\nn\\
 &\hspace{2em}\left.
  +E_6\bigl(-3\tphi_2^2\tphi_8+54\tphi_0\tphi_{12}+1296\tphi_6^2\bigr)
  +E_4^2\bigl(-18\tphi_2\tphi_{12}+72\tphi_6\tphi_8\bigr)
  +E_4E_6\tphi_8^2
\right),\nn\\
\tg_3
 &=-\frac{1}{10077696}
  \left(
   \tphi_2^6
  +54\tphi_0^2\tphi_2^2\tphi_8
  +324\tphi_0\tphi_2^3\tphi_6
  +1944\tphi_0^3\tphi_{12}
  +17496\tphi_0^2\tphi_6^2
  \right.\nn\\
 &\hspace{2em}\left.
  +E_4\bigl(-3\tphi_2^4\tphi_8
   +108\tphi_0^2\tphi_8^2
   -486\tphi_0\tphi_2^2\tphi_{12}
   +1620\tphi_0\tphi_2\tphi_6\tphi_8
   +1944\tphi_2^2\tphi_6^2\bigr)
  \right.\nn\\
 &\hspace{2em}\left.
  +E_6\bigl(
   -36\tphi_0\tphi_2\tphi_8^2
   +90\tphi_2^3\tphi_{12}
   -324\tphi_2^2\tphi_6\tphi_8
   +5832\tphi_0\tphi_6\tphi_{12}
   +46656\tphi_6^3\bigr)
\right.\nn\\
 &\hspace{2em}\left.
  +E_4^2\bigl(3\tphi_2^2\tphi_8^2
   +162\tphi_0\tphi_8\tphi_{12}
   -1944\tphi_2\tphi_6\tphi_{12}
   +3888\tphi_6^2\tphi_8\bigr)
  \right.\nn\\
 &\hspace{2em}\left.
  +E_4E_6\bigl(-54\tphi_2\tphi_8\tphi_{12}+108\tphi_6\tphi_8^2\bigr)
  +E_4^3\bigl(-\tphi_8^3+162\tphi_{12}^2\bigr)
  +E_6^2\bigl(2\tphi_8^3-162\tphi_{12}^2\bigr)
\right).
\label{eq:F4mucoeff}
\end{align}

We observe that the above $\tf_i,\tg_j$ are identical with
the generators of the $D(\mathfrak{d}_4)$-invariant
subring \cite{DelZotto:2017mee}
of $W(D_4)$-invariant Jacobi forms,
where $D(\mathfrak{d}_4)$ denotes the symmetry
group of the $D_4$ Dynkin diagram.
The precise relations are
\begin{align}
\begin{aligned}
\tf_1
 &=-\frac{2}{3}\phi_2^\DGHKKL,\quad&
\tf_2
 &=\frac{4}{3}\phi_8^\DGHKKL,\\
\tg_1
 &=-\frac{1}{18}\phi_0^\DGHKKL,\quad&
\tg_2
 &=\frac{2}{9}\phi_6^\DGHKKL,\quad&
\tg_3
 &=-\frac{8}{27}\phi_{12}^\DGHKKL,
\end{aligned}
\end{align}
where $\phi_{k}^\DGHKKL$ are
given in \cite[eq.(B.33)]{DelZotto:2017mee}.
This is easily checked by using \eqref{eq:F4gen}
to rewrite $\tphi_k=\phi_k(\tau,\mu_j)$
in terms of the $W(D_4)$-invariant
Jacobi forms $\varphi_l(\tau,\mu_j)$, $\psi_4(\tau,\mu_j)$.
The relation between
the $D(\mathfrak{d}_4)$-invariant subring
and the ring $J^{F_4}_{*,*}$ of $W(F_4)$-invariant Jacobi forms
has been discussed in \cite{Duan:2020imo}.
Our result is a good exemplification of this relation.

The fact that two different $W(F_4)$ automorphisms
arise is due to the isomorphism between the
root lattice $L_{D_4}$ and its dual lattice $L_{D_4}^*$.
Specifically, they are given by
\begin{align}
\begin{aligned}
L_{D_4}
 &=\bigl\{\vecom=(\omega_1,\omega_2,\omega_3,\omega_4)\in\bbZ^4
  \bigm| \textstyle\sum_{j=1}^4 \omega_j\in 2\bbZ\bigr\},\\
L_{D_4}^*
 &=\bigl\{\vecom
  \in\bbZ^4\cup\bigl(\bbZ+\tfrac{1}{2}\bigr)^4\bigr\}.
\end{aligned}
\end{align}
These are in fact identical with
the rescaled and unrescaled $F_4$ root lattices respectively
(see \eqref{eq:rescaledLF4}).
The isomorphism between these two lattices is encoded
as the identity between the lattice theta functions:
\begin{align}
\begin{aligned}
\Theta_{D_4^*}(\tau,\vecm)
 &=\sum_{\vecom\in L_{D_4}^*}
   \exp\left(\pi i\tau\vecom^2
   +2\pi i\vecm\cdot\vecom\right)\\
 &=\sum_{\vecom\in L_{D_4}}
   \exp\left(\pi i\tau\frac{\vecom^2}{2}
   +2\pi i\vecmu\cdot\vecom\right)
 =\Theta_{D_4}\left(\frac{\tau}{2},\vecmu\right).
\end{aligned}
\end{align}
From this it is clear that
the two lattices differ only by their scale;
the minimal norms of
$L_{D_4}$ and $L_{D_4}^*$ are 2 and 1 respectively.
The above relation is also expressed as
\begin{align}
T_2(\tau,\vecm)+T_3(\tau,\vecm)
 =
\Theta_{D_4^*}(\tau,\vecm)
 =\Theta_{D_4}\left(\frac{\tau}{2},\vecmu\right)
 =\frac{1}{2}
  \left[T_3\left(\frac{\tau}{2},\vecmu\right)
       +T_4\left(\frac{\tau}{2},\vecmu\right)\right],
\end{align}
where $T_k(\tau,\vecm)$ is given in \eqref{eq:Tdef}.
Recall that $W(D_4)$-invariant Jacobi forms
of index one are certain linear combinations
of $T_k(\tau,\vecm)$ (with some coefficient functions in $\tau$),
while $T_k(\tau/2,\vecmu)$
constitute $W(D_4)$-invariant Jacobi forms of index two.
The same holds for $F_4$. This explains
why the coefficients in the $F_4$ curve \eqref{eq:F4curve}
are Jacobi forms of indices $0,1,2,3$
while the coefficients \eqref{eq:F4mucoeff}
of the curve \eqref{eq:F4mucurve} are Jacobi forms of
indices $0,2,4,6$.

\subsection{5d limit}

\begin{figure}[t]
\begin{center}
\begin{tikzpicture}[scale=0.15]
\coordinate (F1) at ( 0,0);
\coordinate (F2) at (10,0);
\coordinate (F3) at (20,0);
\coordinate (F4) at (30,0);
\draw [thick] (F1) -- (F2);
\draw [thick,double distance=2pt] (F2) -- (F3);
\draw [thick] (F3) -- (F4);
\draw [thick] (14,1) .. controls (14.5,.4) and (15,.2) .. (16,0);
\draw [thick] (14,-1) .. controls (14.5,-.4) and (15,-.2) .. (16,0);
\filldraw [fill=white,thick] (F1) circle [radius=1];
\filldraw [fill=white,thick] (F2) circle [radius=1];
\filldraw [fill=white,thick] (F3) circle [radius=1];
\filldraw [fill=white,thick] (F4) circle [radius=1];
\node at (F1) [below=1.5mm] {$1$};
\node at (F2) [below=1.5mm] {$2$};
\node at (F3) [below=1.5mm] {$3$};
\node at (F4) [below=1.5mm] {$4$};
\coordinate (D1) at (50,8.66);
\coordinate (D2) at (50,-8.66);
\coordinate (D3) at (55,0);
\coordinate (D4) at (65,0);
\draw [thick] (D1) -- (D3);
\draw [thick] (D2) -- (D3);
\draw [thick] (D3) -- (D4);
\filldraw [fill=white,thick] (D1) circle [radius=1];
\filldraw [fill=white,thick] (D2) circle [radius=1];
\filldraw [fill=white,thick] (D3) circle [radius=1];
\filldraw [fill=white,thick] (D4) circle [radius=1];
\node at (D1) [left=1.5mm] {$1$};
\node at (D2) [left=1.5mm] {$2$};
\node at (D3) [left=1.5mm] {$3$};
\node at (D4) [right=1.5mm] {$4$};
\end{tikzpicture}
\caption{Dynkin diagrams for $F_4$ (left) and $D_4$ (right):
numbers attached to nodes denote the labels
of the corresponding fundamental weights.
\label{Fig:Dynkin}}
\end{center}
\end{figure}
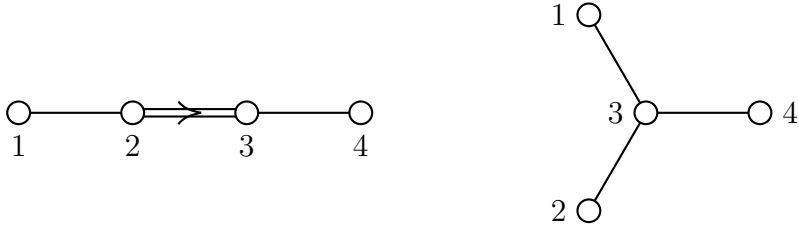

In the limit of $T^2$ shrinking to $S^1$, one obtains
a 5d theory on $\bbR^4\times S^1$.
The corresponding Seiberg--Witten curve is obtained
by merely taking the limit of $q\to 0$.
In this limit, the fundamental $W(F_4)$-invariant
Jacobi forms (including $E_4,E_6$) become
\begin{align}
\begin{aligned}
E_4&=E_6=1,\\
\phi_0&=\frac{2}{3}v_4+32,\\
\phi_2&=2v_4-48,\\
\phi_6&=4v_1-\frac{2}{9}v_4^2+\frac{1}{3}v_3-\frac{4}{3}v_4+32,\\
\phi_8&=4v_4^2-12v_3-48v_4,\\
\phi_{12}
 &=-\frac{8}{9}v_4^3+4v_3v_4-12v_2-32v_4^2+96v_3+384v_4+144v_1+768.
\end{aligned}
\label{eq:phi5d}
\end{align}
Here, $v_j\ (j=1,\ldots,4)$ denote the Weyl orbit characters
associated with the fundamental weights $\vecL_j^{F_4}$ of $F_4$
(see Figure~\ref{Fig:Dynkin}
for our labeling of the fundamental weights)
\begin{align}
v_j(\vecm)
 :=\sum_{\vecv\in\mbox{\scriptsize{Weyl orbit of }}\vecL_j^{F_4}}
   e^{2\pi i\vecv\cdot\vecm}.
\end{align}
Similarly, we define $D_4$ Weyl orbit characters $w_j$ by
\begin{align}
w_j(\vecm)
 :=\sum_{\vecv\in\mbox{\scriptsize{Weyl orbit of }}\vecL_j^{D_4}}
   e^{2\pi i\vecv\cdot\vecm}.
\label{eq:WOCD4}
\end{align}
Then $v_j$ are expressed in terms of $w_k$ as
\begin{align}
\begin{aligned}
v_1 &= w_3,\\
v_2 &= w_1w_2w_4-4\left(w_1^2+w_2^2+w_4^2\right)+12w_3+64,\\
v_3 &= w_1w_2+w_2w_4+w_4w_1-4(w_1+w_2+w_4),\\
v_4 &= w_1+w_2+w_4.
\end{aligned}
\end{align}
Furthermore, $w_j$ are explicitly written as
\begin{align}
\begin{aligned}
w_1&=\sum_\mathrm{even}e^{\pi i(\pm m_1\pm m_2\pm m_3\pm m_4)},\\
w_2&=\sum_\mathrm{odd}e^{\pi i(\pm m_1\pm m_2\pm m_3\pm m_4)},\\
w_3&=\sum_{j<k,\pm,\pm}e^{2\pi i(\pm m_j\pm m_k)},\\
w_4&=\sum_{j,\pm}e^{\pm 2\pi i m_j},
\end{aligned}
\end{align}
where the first (second) sum is taken for all
terms with even (odd) number of `+' in the exponent.

Substituting these into \eqref{eq:F4curve},
one obtains the 5d $F_4$ curve.
Of course,
the same curve is obtained by
evaluating the 5d $E_8$ curve given
in \cite{Minahan:1997ch,Eguchi:2002fc}
at the value \eqref{eq:F4config1}.
In the same manner,
the 5d $F_4$ curve expressed in terms of $\mu_j$ can be obtained
either from \eqref{eq:F4mucurve}--\eqref{eq:F4mucoeff}
or from the 5d $E_8$ curve with \eqref{eq:F4config2}.
As is clear from this construction,
the 5d $F_4$ curve merely describes certain restricted cases
of the 5d $E_8$ theory, namely
the 5d $\mathcal{N}=1\ \grp{SU}(2)$ gauge theory on $\bbR^4\times S^1$
with $N_\mathrm{f}=7$ flavors.

\subsection{4d limit}

In the limit of $T^2$ shrinking to zero size,
one obtains a 4d theory on $\bbR^4$.
It is well known that 
the Seiberg--Witten curve for a 4d theory
is expressed in terms of Casimir invariants.
For a Lie algebra $\alg{g}$,
Casimir invariants form a polynomial algebra
which is the center of the universal enveloping
algebra of $\alg{g}$.
For $F_4$ the polynomial algebra is generated by
four generators $p_2,p_6,p_8,p_{12}$,
where each $p_k$ is a homogeneous polynomial in $m_1,m_2,m_3,m_4$
of degree $k$.
The choice of $p_k$ is not unique.
We define $p_k$ by
\begin{align}
\prod_{\vecal}
 (1-t\vecal\cdot\vecm)
=\sum_{j=0}^{24}p_jt^j,
\end{align}
where the product is taken over all long roots $\vecal$ of $F_4$:
\begin{align}
\vecal\in\{\pm\vece_i\pm\vece_j\,|\,1\le i<j\le 4\}.
\end{align}

In \cite{Eguchi:2002fc} it was clarified
how to obtain the 4d $E_8$ curve \cite{Minahan:1996cj}
from the 5d $E_8$ curve.
Following the method of \cite{Eguchi:2002fc},
let us start with the $F_4$ curve
\eqref{eq:F4curve} with the 5d coefficients \eqref{eq:phi5d}.
We reinstate the circumference $L$ of the $S^1$ in the fifth
dimension by rescaling the variables as
\begin{align}
\begin{aligned}
m_i&\to \frac{L}{2\pi}m_i,\\[1ex]
x&\to L^{10}x,\qquad y\to L^{15}y,\\
u&\to L^6u-p_2 L^2-\frac{p_2^2}{144}L^4
  +\left(\frac{p_6}{480}-\frac{23p_2^3}{103680}\right)L^6.
\end{aligned}
\end{align}
Expanding the curve in $L$,
we see that all the terms up to the order of $L^{29}$ cancel
as in \cite{Eguchi:2002fc}.
At the order of $L^{30}$ we obtain
\begin{align}
\begin{aligned}
y^2=4x^3
 &+\frac{1}{138240}u^2(46080p_2u+1152p_8-528p_2p_6+35p_2^4)x\\
 &+\frac{1}{5374771200}u^3
  \bigl(21499084800u^2+(223948800p_6-21772800p_2^3)u\\[1ex]
 &\hspace{3em}
  +1244160p_{12}
  -85824p_2^2p_8
  +168480p_6^2
  -2424p_2^3p_6
  -245p_2^6
 \bigr).
\end{aligned}
\label{eq:4dF4curve}
\end{align}
We verified that this curve can also be directly obtained from
the original 4d $E_8$ curve of Minahan and Nemeschansky
\cite{Minahan:1996cj}
by setting the mass parameters as in \eqref{eq:F4config1}
and rescaling the variables as
\begin{align}
x\to -4x,\qquad
y\to 4i y,\qquad
\rho\to 2u.
\end{align}

Similarly, we can take the 4d limit of the curve
\eqref{eq:F4mucoeff} expressed in terms of $\vecmu$.
However, the result is simply obtained from \eqref{eq:4dF4curve}
by the change of variables \eqref{eq:F4map}.
Specifically, if we define the generators
$\tp_2,\tp_6,\tp_8,\tp_{12}$ by
\begin{align}
\prod_{\vecal}
 (1-t\vecal\cdot\vecmu)
=\sum_{j=0}^{24}\tp_jt^j,
\end{align}
$p_i$ and $\tp_j$ are related as
\begin{align}
\begin{aligned}
 p_2&=2\tp_2,\\
 p_6&=-8\tp_6+\frac{41}{27}\tp_2^3,\\
 p_8&=16\tp_8-18\tp_2\tp_6+\frac{41}{24}\tp_2^4,\\
 p_{12}&=-64\tp_{12}+\frac{1012}{135}\tp_2^2\tp_8+\frac{38}{3}\tp_6^2
  -\frac{5363}{810}\tp_2^3\tp_6+\frac{4951}{11664}\tp_2^6.
\end{aligned}
\end{align}
By substituting these into \eqref{eq:4dF4curve}
we obtain
\begin{align}
\begin{aligned}
y^2=4x^3
 &+\frac{1}{9720}u^2(6480\tp_2u+1296\tp_8-864\tp_2\tp_6+65\tp_2^4)x\\
 &+\frac{1}{15746400}u^3
  \bigl(62985600u^2+(-5248800\tp_6+486000\tp_2^3)u\\[1ex]
&\hspace{3em}
 -233280\tp_{12}
 +11232\tp_2^2\tp_8
 +77760\tp_6^2
 -17568\tp_2^3\tp_6
 +835\tp_2^6
 \bigr).
\end{aligned}
\label{eq:4dF4curve2}
\end{align}
Physically, the curves \eqref{eq:4dF4curve} and \eqref{eq:4dF4curve2}
merely describe partial mass deformations of the 4d $E_8$ SCFT.

\section{Equivalence with thermodynamic limit of Nekrasov-type formula}
\label{sec:Nek}

In this section we take the thermodynamic limit
of the Nekrasov-type formula \cite{Sakai:2012ik}
for the E-string theory with four Wilson line parameters,
generalizing the result of \cite{Ishii:2013nba}.\footnote{An attempt
in this direction was made earlier in \cite{Ishii:2015iya}.}
In particular, we derive the Seiberg--Witten curve
of the following simple form
\begin{align}
u\left(-Y^2+4X^3-\frac{E_4}{12}X-\frac{E_6}{216}\right)
 +4\prod_{j=1}^4
  \left(\phi_{-2,1}(\tau,\mu_j)X
  -\frac{\phi_{0,1}(\tau,\mu_j)}{12}\right)=0,
\label{eq:thermocurve}
\end{align}
where $\phi_{-2,1},\,\phi_{0,1}$ are the classic weak Jacobi forms
of Eichler and Zagier \cite{EichlerZagier}
\begin{align}
\phi_{-2,1}(\tau,z)
 &=\frac{\varth_1(z,\tau)^2}{\eta(\tau)^6},\qquad
\phi_{0,1}(\tau,z)
  =4\left[
     \frac{\varth_2(z,\tau)^2}{\varth_2(0,\tau)^2}
    +\frac{\varth_3(z,\tau)^2}{\varth_3(0,\tau)^2}
    +\frac{\varth_4(z,\tau)^2}{\varth_4(0,\tau)^2}
    \right].
\label{eq:A1gen}
\end{align}
Moreover, we show that this curve is equivalent to
the $F_4$ curve studied in the last section.

\subsection{Nekrasov-type formula for prepotential of E-string theory}

Let us first briefly recall the Nekrasov-type formula proposed in
\cite{Sakai:2012zq,Sakai:2012ik}.
The formula is obtained as a special case of
the 6d generalization \cite{Hollowood:2003cv}
of the original Nekrasov partition function
for the $\grp{U}(N)$ gauge theory
with $2N$ fundamental matters
\cite{Nekrasov:2002qd}.

Let $\bfR=(R_1,\ldots,R_N)$ denote an $N$-tuple of partitions. Each
partition $R_k$ is a nonincreasing sequence of nonnegative integers
\begin{align}
R_k = \{
  \nu_{k,1}\ge\nu_{k,2}\ge\cdots\ge\nu_{k,\ell(R_k)}>
  \nu_{k,\ell(R_k)+1}=\nu_{k,\ell(R_k)+2}=\cdots=0\}.
\end{align}
Here $\ell(R_k)$ is the number of nonzero $\nu_{k,i}$. $R_k$ is 
represented by a Young diagram. Let $|R_k|$ denote the size of $R_k$,
i.e.~the number of boxes in the Young diagram of $R_k$:
\begin{align}
|R_k| := \sum_{i=1}^\infty\nu_{k,i} = \sum_{i=1}^{\ell(R_k)}\nu_{k,i}.
\end{align}
Similarly, the size of $\bfR$ is denoted by
\begin{align}
|\bfR| := \sum_{k=1}^N |R_k|.
\end{align}
Let $\Rt_k=\{\nu_{k,1}^\vee\ge\nu_{k,2}^\vee\ge\cdots\}$ denote
the conjugate partition of $R_k$.
We also introduce the notation
\begin{align}
h_{k,l}(i,j):=\nu_{k,i}+\nu_{l,j}^\vee-i-j+1,
\end{align}
which represents the relative hook-length of a box at $(i,j)$
between the Young diagrams of $R_k$ and $R_l$.

In \cite{Hollowood:2003cv}, Hollowood, Iqbal and Vafa proposed
the 6d (or elliptic) generalization of
the original Nekrasov partition function for
the $\grp{U}(N)$ gauge theory with $2N$ fundamental matters
\cite{Nekrasov:2002qd}. It takes the simple form:
\begin{align}
Z:=\sum_{\bfR}
\left(-e^{2\pi i\varphi}\right)^{|\bfR|}
\prod_{k=1}^N
\prod_{(i,j)\in R_k}
\frac
{\prod_{n=1}^{2N}
 \varth_1\left(a_k-m_n+\tfrac{1}{2\pi}(j-i)\hbar,\tau\right)}
{\prod_{l=1}^N
\varth_1\left(a_k-a_l+\tfrac{1}{2\pi}h_{k,l}(i,j)\hbar,\tau\right)^2}.
\label{Zconformal}
\end{align}
Here the sum is taken over all possible partitions $\bfR$,
including the empty partition. The indices $(i,j)$ run over
the coordinates of all boxes in the Young diagram of $R_k$. Note that
the consistency condition
\begin{align}
2\sum_{k=1}^N a_k - \sum_{n=1}^{2N} m_n = 0
\label{amrel}
\end{align}
is required, where the equality is understood modulo
periods of the torus $\bbC/(\bbZ+\tau\bbZ)$.

Let $\omega_k\ (k=1,2,3,4)$ be half periods of the torus:
\begin{align}
\omega_1=\frac{1}{2},\qquad
\omega_2=-\frac{1+\tau}{2},\qquad
\omega_3=\frac{\tau}{2},\qquad
\omega_4=0.
\label{eq:halfperiods}
\end{align}
The Nekrasov-type formula \cite{Sakai:2012zq,Sakai:2012ik}
for the E-string theory
with Wilson line parameters \eqref{eq:F4config2}
is obtained by simply setting
\begin{align}
N=4,\qquad
a_k=\omega_k\quad(k=1,2,3,4),\qquad
m_n=-m_{n+4}=\mu_n\quad(n=1,2,3,4).
\label{fourWsetup}
\end{align}
The Seiberg--Witten prepotential for the E-string theory
is then given by
\begin{align}
F_0 = \left(2\hbar^2\ln Z\right)\big|_{\hbar=0}\,.
\label{F0inZ}
\end{align}
An important remark is that
the above Nekrasov-type formula is limited to the genus zero
prepotential and does not work for higher genus parts
\cite{Sakai:2012zq}.
This is in contrast to the ordinary Nekrasov partition functions.
However, this does not cause any problem in deriving
the Seiberg--Witten curve since only the genus zero part concerns.

The fact that the E-string prepotential is extracted
from the Nekrasov partition function for the $\grp{U}(4)$ gauge theory
is somewhat surprising, since there is currently no known duality nor
renormalization group flow that directly relates the two theories.
It could be just a mathematical coincidence,
but in the rest of this subsection let us
elaborate on the underlying physics
which may support this relation.

Field theories under consideration have
6d (1,0) supersymmetry and are thus chiral theories.
For a 6d (1,0) gauge theory without gravity,
only a few specific matter contents are allowed
in order for the theory to be free of anomaly
\cite{Seiberg:1996qx, Danielsson:1997kt}.
The $\grp{U}(N)$ gauge theory with $2N$ fundamental matters
is one of such allowed examples; in this case
the anomaly can be canceled via Green--Schwarz mechanism
by adding a tensor multiplet to the theory.
But a single tensor multiplet
is exactly the supermultiplet that the E-strings couple to.
In other words, the above gauge theory becomes anomaly-free
if it is coupled to the E-string theory.
It should be possible, in principle, to derive
the Nekrasov partition function \eqref{Zconformal}
from the path integral that involves a tensor multiplet.
Therefore, it would not be surprising that
the Nekrasov partition function \eqref{Zconformal}
knows about the spectrum of E-strings.

There are some supporting evidence for the above argument.
First, the above discussion does not require $N$ to be $4$.
This is indeed the case.
In this paper we consider the case of $N=4$,
but the E-string prepotential can also be obtained
from the $N=3$ case when the global $E_8$ symmetry is unbroken
\cite{Sakai:2012zq,Sakai:2012ik}
and from the $N=2$ case
when the global symmetry is broken to, for example, $E_7\oplus A_1$
\cite{Sakai:2012ik}.
Second, in the 6d (1,0) gauge theory with a tensor multiplet,
the scalar $\Phi$ in the tensor multiplet
is coupled to the gauge field $F_{\mu\nu}$
through the bosonic term $\Phi F_{\mu\nu}^2$
in the Lagrangian \cite{Seiberg:1996qx}.
This naturally explains why the tension $\varphi$ of E-strings,
which is given by the expectation value of $\Phi$,
appears in \eqref{Zconformal}
as the chemical potential for the instanton number.

In the following subsections we will give a technically solid proof
of the Nekrasov-type formula \eqref{F0inZ},
but it would be very interesting to substantiate (or disprove)
the above argument and clarify the physical origin of
the formula.

\subsection{Thermodynamic limit of Nekrasov-type formula}

Nekrasov and Okounkov proved that
the Seiberg--Witten curve is obtained from
the Nekrasov partition function by taking the thermodynamic limit
\cite{Nekrasov:2003rj}.
In \cite{Ishii:2013nba} the thermodynamic limit of the above
Nekrasov-type formula was studied for the massless case $\mu_j=0$
and also for some other special cases.
Since the derivation is rather technical and all the fine points
were explained in full detail in \cite{Ishii:2013nba},
here we only describe how the derivation in \cite{Ishii:2013nba}
is modified for the case of general values of $\mu_j$.

In the rank-one Seiberg--Witten theory, 
the expectation value $\varphi$ of the Higgs field is expressed
through the relation
\cite{Seiberg:1994rs,Seiberg:1994aj}
\begin{align}
\frac{\partial\varphi}{\partial u}
 =\frac{i}{4\pi^2}\oint_{\alpha}\frac{dx}{y}
 =\frac{i}{4\pi^2u}\oint_{\alpha}\frac{dX}{Y}.
\label{eq:dphiinxy}
\end{align}
Here $\alpha$ is a fundamental one-cycle of the torus
and we have introduced the rescaled variables
$X=u^{-2}x,Y=u^{-3}y$.
On the other hand, by taking the thermodynamic limit of
the Nekrasov-type formula,
the same quantity is expressed as \cite{Ishii:2013nba}
\begin{align}
\frac{\partial\varphi}{\partial u}
 =\frac{i}{2\pi u}\oint_{\alpha}\frac{dz}{\sqrt{1-H(z)^{-1}}}.
\label{eq:dphiinH}
\end{align}
Here function $H(z)$ for the case of
Wilson line configuration \eqref{eq:F4config2}
is given by \cite{Ishii:2013nba}
\begin{align}
H(z)=\kappa
 \frac{\prod_{j=1}^4\varth_1(z-\omega_j,\tau)^2}
      {\prod_{n=1}^8\varth_1(z-m_n,\tau)}
\end{align}
with
\begin{align}
\kappa=\frac{q^{1/2}\eta^{12}}{4}u.
\end{align}
$\omega_j$ and $m_n$ are given in
\eqref{eq:halfperiods} and \eqref{fourWsetup} respectively.
By using the identities
\begin{align}
\begin{aligned}
\prod_{j=1}^3\varth_1(z-\omega_j,\tau)
 &=-\prod_{j=1}^3\varth_1(z+\omega_j,\tau),\\
\varth_1(z+w,\tau)\varth_1(z-w,\tau)
 &=-(2\pi)^{-2}\eta^{-6}\varth_1(z,\tau)^2\varth_1(w,\tau)^2
   \left[\wp(z)-\wp(w)\right],\\
\prod_{j=1}^3\varth_1(\omega_j,\tau)^2&=-4q^{-1/2}\eta^6,\\[-2ex]
\wp'(z)^2
 &=4\prod_{j=1}^3\left[\wp(z)-\wp(\omega_j)\right],
\end{aligned}
\end{align}
$H(z)$ is rewritten as
\begin{align}
\begin{aligned}
H(z)&=\pi^2q^{1/2}\eta^{18}u
 \frac{\prod_{j=1}^3\left(\varth_1(\omega_j,\tau)^2
         \left[\wp(z)-\wp(\omega_j)\right]\right)}
      {\prod_{j=1}^4\left(\varth_1(\mu_j,\tau)^2
         \left[\wp(z)-\wp(\mu_j)\right]\right)}\\
&=-\pi^2\eta^{24}u
 \frac{\wp'(z)^2}
      {\prod_{j=1}^4\left(\varth_1(\mu_j,\tau)^2
         \left[\wp(z)-\wp(\mu_j)\right]\right)}.
\end{aligned}
\end{align}
\eqref{eq:dphiinH} is then written as
\begin{align}
\frac{\partial\varphi}{\partial u}
 =\frac{i}{2\pi u}\oint_{\alpha}
  \frac{\wp'(z)dz}
       {\sqrt{\wp'(z)^2+\pi^{-2}\eta^{-24}u^{-1}
        \prod_{j=1}^4\left(\varth_1(\mu_j,\tau)^2
        \left[\wp(z)-\wp(\mu_j)\right]\right)}}.
\end{align}
This is identified with \eqref{eq:dphiinxy}
by the map
\begin{align}
\begin{aligned}
X&=(2\pi)^{-2}\wp(z),\\
Y&=(2\pi)^{-3}
\sqrt{\wp'(z)^2+\pi^{-2}\eta^{-24}u^{-1}
        \textstyle\prod_{j=1}^4\left(\varth_1(\mu_j,\tau)^2
        \left[\wp(z)-\wp(\mu_j)\right]\right)}.
\end{aligned}
\end{align}
Substituting the identity
\begin{align}
\wp'(z)^2
 =4\wp(z)^3-\frac{(2\pi)^4}{12}E_4\wp(z)-\frac{(2\pi)^6}{216}E_6
\end{align}
and eliminating $\wp(z)$, one obtains
\begin{align}
Y^2=4X^3-\frac{E_4}{12} X-\frac{E_6}{216}
+4u^{-1}\prod_{j=1}^4\left(\frac{\varth_1(\mu_j,\tau)^2}{\eta^6}
 \left[X-\frac{\wp(\mu_j)}{(2\pi)^2}\right]\right).
\label{eq:thermopre}
\end{align}
Finally using the well-known relation
\begin{align}
\phi_{0,1}(\tau,z)
=\frac{12}{(2\pi)^2}\phi_{-2,1}(\tau,z)\wp(z),
\end{align}
one obtains \eqref{eq:thermocurve}.

We close this subsection with a few remarks
in connection with some related works.
First, in the limit of $\mu_4\to 0$,
the quartic curve \eqref{eq:thermocurve} reduces to the cubic one
\begin{align}
u\left(-Y^2+4X^3-\frac{E_4}{12}X-\frac{E_6}{216}\right)
 -4\prod_{j=1}^3
  \left(\phi_{-2,1}(\tau,\mu_j)X
  -\frac{\phi_{0,1}(\tau,\mu_j)}{12}\right)=0.
\end{align}
Essentially the same curve
(expressed in terms of $\wp(\mu_j)$ as in \eqref{eq:thermopre})
was considered in \cite{Mohri:2001zz}.
Our curve \eqref{eq:thermocurve} can be viewed as
its generalization to the case of four Wilson line parameters.
Second, instead of using the variables $X$ and $Y$,
one can write \eqref{eq:dphiinH} as
\begin{align}
\frac{\partial\varphi}{\partial u}
 =\frac{i}{2\pi u}\oint_{\alpha}\frac{dz}{w},\qquad
w=\sqrt{1-H(z)^{-1}}
\end{align}
and regard the second equation
as the Seiberg--Witten curve expressed
in the variables $z$ and $w$.
By using the identity
\begin{align}
\prod_{j=1}^4\varth_1(z-\omega_j,\tau)
 =iq^{-1/4}\eta^{3}\varth_1(2z,\tau),
\end{align}
this curve can be rewritten as
\begin{align}
 u (w^2-1)\eta^{18}\varth_1(2z,\tau)^2
 =4\prod_{j=1}^4[\varth_1(z+\mu_j,\tau)\varth_1(z-\mu_j,\tau)].
\label{eq:nonellF4}
\end{align}
The curve of this form has been
discussed in \cite{Haghighat:2018dwe,Chen:2021ivd}.

\subsection{Equivalence}

We are now in a position to show that the curve \eqref{eq:thermocurve}
is equivalent to the $F_4$ curve studied
in the last section. To do this, we first rescale the variables
by the replacement\footnote{$X,Y$ after the rescaling are
identical with the original $x,y$ appeared in \eqref{eq:dphiinxy};
here we use the letters $X,Y$ for the variables of the quartic curve
and save $x,y$ for the cubic curve \eqref{eq:cubic}.}
$X\to u^{-2}X,\ Y\to u^{-3}Y$
and write the curve \eqref{eq:thermocurve} as
\begin{align}
Y^2=4X^3-\frac{E_4}{12}u^4X-\frac{E_6}{216}u^6
 +4u^5\prod_{j=1}^4
  \left(\phi_{-2,1}(\mu_j)\frac{X}{u^2}
 -\frac{\phi_{0,1}(\mu_j)}{12}\right).
\label{eq:thermocurve2}
\end{align}
This is a quartic curve of the form
\begin{align}
Y^2=c_0X^4+c_1X^3+c_2X^2+c_3X+c_4.
\label{eq:quartic}
\end{align}
It is well known that any quartic curve of this form 
with nonzero $c_0$ can be transformed into the Weierstrass form
$y^2=4x^3-fx-g$: By the change of
variables $Y\to c_0^{1/2}Y,\ X\to X-c_1/(4c_0)$,
\eqref{eq:quartic} is rewritten in the form
\begin{align}
Y^2=X^4+6\tc_2 X^2+4\tc_3 X+\tc_4,
\end{align}
which is converted to the Weierstrass form by the map
\begin{align}
2X(x+\tc_2)=y-\tc_3,\qquad Y=2x-X^2-\tc_2.
\end{align}
By further rescaling the variables as
$x\to c_0^{-1}x,\ y\to c_0^{-3/2}y$,
the quartic curve \eqref{eq:quartic} finally becomes
\begin{align}
y^2=4x^3-\left(c_0c_4-\frac{c_1c_3}{4}+\frac{c_2^2}{12}\right)x
-\left(\frac{c_0c_2c_4}{6}-\frac{c_0c_3^2}{16}+\frac{c_1c_2c_3}{48}
  -\frac{c_1^2c_4}{16}-\frac{c_2^3}{216}\right).
\label{eq:cubic}
\end{align}
It can be easily checked that the curves \eqref{eq:quartic} and 
\eqref{eq:cubic} have the same discriminant up to the normalization.
In this way, one can calculate a Weierstrass elliptic curve
equivalent to \eqref{eq:thermocurve2}.
By direct calculation 
we verified that this precisely gives
the $F_4$ curve \eqref{eq:F4mucurve}--\eqref{eq:F4mucoeff},
where $u$ in \eqref{eq:thermocurve2} is identified
with $\tu$ in \eqref{eq:F4mucurve}.

\section{$D_4$ triality invariant model}\label{sec:D4tri}

\subsection{Emergence of $D_4$ triality in E-string theory}

One of the purposes of this paper is to establish a natural
flow from the 6d E-string theory to
the 4d $\mathcal{N}=2$ supersymmetric $\grp{SU(2)}$ gauge theory
with four massive flavors ($N_\mathrm{f}=4$ theory).
It is sometimes believed that
the E-string theory with half of the Wilson line parameters
turned on as in \eqref{eq:F4config1}
gives a natural 6d generalization of the $N_\mathrm{f}=4$ theory,
since the system exhibits a $W(D_4)$ automorphism
acting on $m_i$ (which is actually promoted to $W(F_4)$ as we have
seen) and keeps a global $D_4$ symmetry
(acting on four $0$'s in \eqref{eq:F4config1}).
However, this is not the case:\footnote{As mentioned in
section~\ref{sec:intro},
there is a way to identify this 6d setup with
the 4d $N_\mathrm{f}=4$ theory
at the price of losing natural identification of mass parameters
\cite{Sakai:2012ik}.}
In the limit of $m_i\to 0$,
the 6d theory specified by \eqref{eq:F4config1}
recovers the global $E_8$ symmetry
and correspondingly the elliptic fibration described by
the Seiberg--Witten curve
exhibits a singular fiber of Kodaira type $\mathrm{II}^*$,
which corresponds to an $E_8$ singularity.
On the other hand, the elliptic fibration of
the $N_\mathrm{f}=4$ theory in the massless limit
exhibits two singular fibers of Kodaira type $\mathrm{I}^*_0$,
which correspond to two $D_4$ singularities.

In the massless case,
a natural 6d generalization of the $N_\mathrm{f}=4$ theory
was studied in detail in \cite{Sakai:2014hsa}.
The corresponding Wilson line parameters are given by
\begin{align}
\vecm
 =(0,\tfrac{1}{2},-\tfrac{1+\tau}{2},\tfrac{\tau}{2},
   0,\tfrac{1}{2},-\tfrac{1+\tau}{2},\tfrac{\tau}{2}).
\end{align}
Note that by using the $W(E_8)$ action \eqref{eq:vecmu2m}
this is mapped to
\begin{align}
\vecm=(
 \tfrac{1}{2},-\tfrac{1}{2},-\tfrac{1}{2},-\tfrac{1}{2}-\tau,0,0,0,0).
\end{align}
The Wilson line setup for
any natural 6d generalization of the $N_\mathrm{f}=4$ theory
should reduce to a configuration equivalent to these
in the limit of $m_i\to 0$.

A natural candidate is
\begin{align}
\vecm
 =\vecm_\tri
 :=(m_1,m_2,m_3,m_4,
 \tfrac{1}{2},-\tfrac{1}{2},-\tfrac{1}{2},-\tfrac{1}{2}-\tau).
\label{eq:trconf1}
\end{align}
By the discussion in section~\ref{sec:F4auto},
it is clear that this configuration exhibits
$W(D_4)$ automorphism acting on $m_1,\ldots,m_4$.
In what follows we will show that this configuration
in addition exhibits extra automorphisms, namely the $D_4$ triality
combined with modular transformations.

First, as in \eqref{eq:F4refl},
consider the action of $s_{\vece_4}$ on $\vecm_\tri$:
\begin{align}
s_{\vece_4}(\vecm_\tri)
=(m_1,m_2,m_3,-m_4,
 \tfrac{1}{2},-\tfrac{1}{2},-\tfrac{1}{2},-\tfrac{1}{2}-\tau).
\end{align}
This is not equivalent to $\vecm_\tri$,
meaning that the theory is no longer $W(F_4)$-invariant.
However, if one combines $s_{\vece_4}$ with
the modular transformation $\tau\to \tau+1$,
one obtains
\begin{align}
(m_1,m_2,m_3,-m_4,
 \tfrac{1}{2},-\tfrac{1}{2},-\tfrac{1}{2},-\tfrac{1}{2}-\tau-1).
\end{align}
By the $W(E_8)$ action $s_{\vece_4-\vece_5}s_{\vece_4+\vece_5}$,
this is mapped to
\begin{align}
\begin{aligned}
&(m_1,m_2,m_3,m_4,
 -\tfrac{1}{2},-\tfrac{1}{2},-\tfrac{1}{2},-\tfrac{1}{2}-\tau-1)\\
&=\vecm_\tri-(0,0,0,0,1,0,0,1).
\end{aligned}
\end{align}
By the periodicity condition \eqref{eq:periodicity},
this is equivalent to $\vecm_\tri$.

Next, consider
\begin{align}
\begin{aligned}
&s_{\frac{1}{2}(\vece_1-\vece_2-\vece_3+\vece_4)}(\vecm_\tri)\\
&=
(\tfrac{m_1+m_2+m_3-m_4}{2},
 \tfrac{m_1+m_2-m_3+m_4}{2},
 \tfrac{m_1-m_2+m_3+m_4}{2},
 \tfrac{-m_1+m_2+m_3+m_4}{2},
 \tfrac{1}{2},-\tfrac{1}{2},-\tfrac{1}{2},-\tfrac{1}{2}-\tau).
\end{aligned}
\end{align}
This is again not equivalent to $\vecm_\tri$.
However, further applying 
the modular transformation $m_i\to m_i/\tau$, $\tau\to -1/\tau$,
one obtains
\begin{align}
(\tfrac{m_1+m_2+m_3-m_4}{2\tau},
 \tfrac{m_1+m_2-m_3+m_4}{2\tau},
 \tfrac{m_1-m_2+m_3+m_4}{2\tau},
 \tfrac{-m_1+m_2+m_3+m_4}{2\tau},
 \tfrac{1}{2},-\tfrac{1}{2},-\tfrac{1}{2},
 -\tfrac{1}{2}+\tfrac{1}{\tau}).
\end{align}
By the S-transformation
(which is a part of the modular invariance \eqref{eq:SL2Zinv})
of the original E-string theory
this is equivalent to
\begin{align}
(\tfrac{m_1+m_2+m_3-m_4}{2},
 \tfrac{m_1+m_2-m_3+m_4}{2},
 \tfrac{m_1-m_2+m_3+m_4}{2},
 \tfrac{-m_1+m_2+m_3+m_4}{2},
 \tfrac{\tau}{2},-\tfrac{\tau}{2},
-\tfrac{\tau}{2},-\tfrac{\tau}{2}+1).
\label{eq:trconfmid}
\end{align}
Next, applying the $W(E_8)$ action
\begin{align}
 s_{\vece_5+\vece_8}s_{\vece_2-\vece_3}s_{\vece_1-\vece_4}s_\vecs
 s_{\vece_3-\vece_4}s_{\vece_3+\vece_4}
 s_{\vece_1-\vece_2}s_{\vece_1+\vece_2}s_\vecs
\end{align}
on \eqref{eq:trconfmid}, one obtains
\begin{align}
\begin{aligned}
&(m_1,m_2,m_3,m_4,
 -\tfrac{1}{2},-\tfrac{1}{2},-\tfrac{1}{2},\tfrac{1}{2}-\tau)\\
&=\vecm_\tri+(0,0,0,0,-1,0,0,1).
\end{aligned}
\end{align}
Again, by the periodicity condition \eqref{eq:periodicity}
this is equivalent to $\vecm_\tri$.
Note that $\tau$ is also mapped to itself by the whole sequence
of transformations.

Thus we have shown that the system specified by $\vecm=\vecm_\tri$
is invariant under not only $W(D_4)$, but also
\begin{align}
\begin{aligned}
s_{\vece_4}\quad&\mbox{with}\quad\tau\to\tau+1,\\
s_{\frac{1}{2}(\vece_1-\vece_2-\vece_3+\vece_4)}
\quad&\mbox{with}\quad
(\tau,m_1,\ldots,m_4)\to
\left(-\frac{1}{\tau},\frac{m_1}{\tau},\ldots,\frac{m_4}{\tau}\right).
\end{aligned}
\label{eq:TSmtri}
\end{align}
As we will see in the next subsection,
$W(D_4)$ and the above transformations generate
an automorphism group, which was originally found as a symmetry of
the $N_\mathrm{f}=4$ theory \cite{Seiberg:1994aj}.

The configuration $\vecm=\vecm_\tri$ in \eqref{eq:trconf1}
is not the only one that exhibits the $D_4$ triality,
but there are several equivalent expressions.
For instance, the configuration
\begin{align}
\vecm
=(m_1+\tfrac{1}{2},
  m_2-\tfrac{1}{2},
  m_3-\tfrac{1}{2},
  m_4-\tfrac{1}{2}-\tau,0,0,0,0)
\label{eq:trconf2}
\end{align}
is equivalent to $\vecm_\tri$.
This can be seen immediately by rewriting it as
\begin{align}
\begin{aligned}
&\hspace{-1em}
 (m_1+\tfrac{1}{2},
  m_2-\tfrac{1}{2},
  m_3-\tfrac{1}{2},
  m_4-\tfrac{1}{2}-\tau,0,0,0,0)\\
&=
\vecm_\tri
 +(\tfrac{1}{2},-\tfrac{1}{2},-\tfrac{1}{2},-\tfrac{1}{2},
   -\tfrac{1}{2},\tfrac{1}{2},\tfrac{1}{2},\tfrac{1}{2})
 +\tau(0,0,0,-1,0,0,0,1).
\end{aligned}
\label{eq:trconf1-2rel}
\end{align}
The equivalence follows from
the periodicity condition \eqref{eq:periodicity}.

Another useful expression is given in terms of $\mu_i$.
By using the map \eqref{eq:vecmu2m} inversely
with the identification \eqref{eq:F4map},
the configuration \eqref{eq:trconf2} is mapped to
\begin{align}
\begin{aligned}
\vecm
&=(\mu_1,\mu_2+\tfrac{1}{2},
   \mu_3-\tfrac{1+\tau}{2},\mu_4+\tfrac{\tau}{2},
   \mu_1,\mu_2+\tfrac{1}{2},
   \mu_3-\tfrac{1+\tau}{2},\mu_4+\tfrac{\tau}{2}).
\end{aligned}
\label{eq:trconf3}
\end{align}

To sum up,
the $D_4$ triality invariant configuration is obtained from
the $F_4$ configuration studied in section~\ref{sec:F4}
by either adding
$(\tfrac{1}{2},-\tfrac{1}{2},-\tfrac{1}{2},-\tfrac{1}{2}-\tau)$
to $(m_1,\ldots,m_4)$
or adding the half periods
$(0,\tfrac{1}{2},-\tfrac{1+\tau}{2},\tfrac{\tau}{2})$
to $(\mu_1,\ldots,\mu_4)$.

\subsection{$D_4$ triality invariant Jacobi forms}\label{sec:trJacobi}

We have seen that the E-string theory with
the configuration \eqref{eq:trconf1} exhibits
a set of automorphisms which involves both $W(D_4)$
and the $D_4$ triality combined with modular transformations.
The automorphism of this kind was first found
by Seiberg and Witten in the study of
4d $N_\mathrm{f}=4$ theory \cite{Seiberg:1994aj}.
The whole automorphism forms a group,
which we call the modular triality group.
In this subsection we clarify the structure of
this group and introduce
Jacobi forms which respect this automorphism.

Let us first define the modular triality group.
Let $(g,w)$ denote an element of the direct product
$\grp{SL}(2,\bbZ)\times W(F_4)$ which acts on $\bbH\times\bbC^4$ as
\begin{align}
(g,w):(\tau,\vecz)\mapsto
\left(\frac{a\tau+b}{c\tau+d},\frac{w(\vecz)}{c\tau+d}\right),
\quad\mbox{where}\quad
g=\left(\begin{array}{cc}a&b\\ c&d\end{array}\right).
\end{align}
We call the subgroup $\Gtri$ generated by the following elements
the modular triality group:
\renewcommand{\theenumi}{\roman{enumi}}
\renewcommand{\labelenumi}{(\theenumi)}
\begin{enumerate}
\item $(g,1)$ with $g\in \Gamma(2)\subset \grp{SL}(2,\bbZ)$.

\item $(1,w)$ with $w\in W(D_4)\subset W(F_4)$.

\item $\cT=(g_\cT,w_\cT)$ with
\begin{align}
\begin{aligned}
g_\cT:=\left(\begin{array}{cc}1&1\\ 0&1\end{array}\right),
\qquad
w_\cT
\left(\begin{array}{c}z_1\\ z_2\\ z_3\\ z_4\end{array}\right)
:=
\left(\begin{array}{cccc}
1&0&0&0\\
0&1&0&0\\
0&0&1&0\\
0&0&0&-1\end{array}\right)
\left(\begin{array}{c}z_1\\ z_2\\ z_3\\ z_4\end{array}\right).
\end{aligned}
\end{align}

\item $\cS=(g_\cS,w_\cS)$ with
\begin{align}
\begin{aligned}
g_\cS:=\left(\begin{array}{cc}0&-1\\ 1&0\end{array}\right),
\qquad
w_\cS
\left(\begin{array}{c}z_1\\ z_2\\ z_3\\ z_4\end{array}\right)
:=
\frac{1}{2}
\left(\begin{array}{cccc}
1&1&1&-1\\
1&1&-1&1\\
1&-1&1&1\\
-1&1&1&1\end{array}\right)
\left(\begin{array}{c}z_1\\ z_2\\ z_3\\ z_4\end{array}\right).
\end{aligned}
\end{align}
\end{enumerate}
Here the principal congruence subgroup of level $N$ is defined as
\begin{align}
\Gamma(N)=
\left\{
\left(\begin{array}{cc}a&b\\ c&d\end{array}\right)
\in\grp{SL}(2,\bbZ)\biggm|
a\equiv d\equiv 1\ (\text{mod}\ N),\ b\equiv c\equiv 0\ (\text{mod}\ N)
\right\}.
\end{align}
As clearly seen, $\cT$ and $\cS$
represent the transformations in \eqref{eq:TSmtri}
under the identification $z_i=m_i$.
It is also clear that $\Gtri$ satisfies the following inclusions:
\begin{align}
\Gamma(2)\times W(D_4)
\ \subset\  \Gtri
\ \subset\
\grp{SL}(2,\bbZ)\times W(F_4).
\label{eq:incl}
\end{align}

$\Gtri$ can also be described
more abstractly as follows:
Let $f_1,f_2$ be group homomorphisms
given by the quotient maps
\begin{align}
\begin{aligned}
f_1&:\grp{SL}(2,\bbZ)\to\grp{SL}(2,\bbZ)/\Gamma(2)\cong S_3,\\
f_2&:W(F_4)\to W(F_4)/W(D_4)\cong S_3.
\end{aligned}
\end{align}
We identify the two $S_3$'s in such a way that
the correspondence of the generators is given by
$g_\cT\leftrightarrow w_\cT,\,g_\cS\leftrightarrow w_\cS$.
Then we can define $\Gtri$ as the fiber product
of $\grp{SL}(2,\bbZ)$ and $W(F_4)$ over the common $S_3$:
\begin{align}
\Gtri=\grp{SL}(2,\bbZ)\times_{S_3}W(F_4)
 =\{(g,w)\in\grp{SL}(2,\bbZ)\times W(F_4)\,|\,f_1(g)=f_2(w)\}.
\end{align}

Let us next define $\Gtri$-invariant Jacobi forms,
which we call the triality invariant Jacobi forms.
Let $k,m$ be integers $(m\ge 0)$
and $\Gtri$ the modular triality group defined above. We call
a holomorphic function $\varphi_{k,m}:\bbH\times\bbC^4\to \bbC$
a triality invariant weak Jacobi form
of weight $k$ and index $m$
if it satisfies the following properties:
\renewcommand{\theenumi}{\roman{enumi}}
\renewcommand{\labelenumi}{(\theenumi)}
\begin{enumerate}
\item Weyl invariance and modular transformation law:
\begin{align}
&
\varphi_{k,m}\left(
\frac{a\tau+b}{c\tau+d}\,,\frac{w(\vecz)}{c\tau+d}\right)
=(c\tau+d)^k\exp\left(m\pi i\frac{c}{c\tau+d}\,\vecz^2\right)
\varphi_{k,m}(\tau,\vecz),\\[1ex]
&
\left(\left(\begin{array}{cc}a&b\\ c&d\end{array}\right),w\right)
\in \Gtri.\nn
\end{align}

\item Quasi-periodicity:
\begin{align}
\varphi_{k,m}(\tau,\vecz+\tau\vecal+\vecbe)
=e^{-m \pi i (\tau\vecal^2+2\vecz\cdot\vecal)}
\varphi_{k,m}(\tau,\vecz),\qquad
\vecal,\vecbe\in L_{D_4}.
\end{align}

\item $\varphi_{k,m}(\tau,\vecz)$ admits
a Fourier expansion of the form
\begin{align}
\varphi_{k,m}(\tau,\vecz)
=\sum_{n=0}^\infty
 \sum_{\vecw\in L_{D_4}^*}
 c(n,\vecw)e^{2\pi i\vecw\cdot\vecz}q^{n/2}.
\end{align}
\end{enumerate}
It follows immediately that triality invariant Jacobi forms
of index 0 are ordinary modular forms.

Let
\begin{align}
J^\Gtri_{k,m}
\end{align}
denote the vector space of
triality invariant Jacobi forms of weight $k$ and index $m$.
As in the case of $W(R)$-invariant Jacobi forms,
the space of all triality invariant Jacobi forms
\begin{align}
J^\Gtri_{*,*}
 :=\bigoplus_{k\in\bbZ,\,m\in\bbZ_{\ge 0}}J^\Gtri_{k,m}
\end{align}
is a bigraded algebra over the ring of modular forms.
Let $J^{D_4\Gamma(2)}_{*,*}$ denote the bigraded algebra
of $W(D_4)$-invariant Jacobi forms over the ring of modular forms
for $\Gamma(2)$.
By the inclusions \eqref{eq:incl}
we have
\begin{align}
J^{F_4}_{*,*}\subset
J^{\Gtri}_{*,*}\subset
J^{D_4\Gamma(2)}_{*,*}.
\end{align}
This in particular means that
$W(F_4)$-invariant Jacobi forms are automatically
triality invariant Jacobi forms.

Moreover, there exist triality invariant Jacobi forms
that are not $W(F_4)$-invariant.
For instance, one can verify that the following combinations of
$W(D_4)$-invariant Jacobi forms $\varphi_4,\psi_4$
give such triality invariant Jacobi forms:
\begin{align}
\begin{aligned}
\chi_2
 &=(\varth_3^4+\varth_4^4)\varphi_4+3\varth_2^4\psi_4,\\
\chi_0
 &=(2E_4-3\varth_2^8)\varphi_4-3(\varth_3^8-\varth_4^8)\psi_4.
\end{aligned}
\end{align}
Investigating the structure of 
the ring $J^{\Gtri}_{*,*}$ of triality invariant Jacobi forms
is an interesting problem.
It turns out that $J^{\Gtri}_{*,*}$ is not a polynomial algebra,
as in the case of $J^{E_8}_{*,*}$.
The details will be studied separately
in a forthcoming paper \cite{Sakai:2023}.
Here we only mention that any triality invariant Jacobi form
is expressed in terms of the following five fundamental Jacobi forms
\begin{align}
\phi_0\in J^\Gtri_{0,1},\quad
\chi_0\in J^\Gtri_{0,1},\quad
\phi_2\in J^\Gtri_{-2,1},\quad
\chi_2\in J^\Gtri_{-2,1},\quad
\phi_6\in J^\Gtri_{-6,2},
\end{align}
where $\phi_k$ are the $W(F_4)$-invariant Jacobi forms
given in \eqref{eq:F4gen}.
To be more exact, one can prove that
any triality invariant Jacobi form of index $m$
is expressed uniquely as \cite{Sakai:2023}
\begin{align}
\frac{P(E_4,E_6,\phi_0,\chi_0,\phi_2,\chi_2,\phi_6)}
     {\Delta^{[m/2]}},
\end{align}
where $P$ denotes a polynomial of the given variables
and $[m/2]$ is the integer part of $m/2$.
Indeed, $W(F_4)$-invariant Jacobi forms $\phi_8,\phi_{12}$
are expressed as
\begin{align}
\phi_8&=
 \frac{E_4\chi_0^2-2E_6\chi_0\chi_2+E_4^2\chi_2^2}{6912\Delta},\qquad
\phi_{12}=
 \frac{\chi_0^3-3E_4\chi_0\chi_2^2+2E_6\chi_2^3}{124416\Delta}.
\end{align}
%

\subsection{$D_4$ triality invariant curve}

In this subsection we determine the Seiberg--Witten curve
for the E-string theory at the value \eqref{eq:trconf1},
which we call the triality invariant curve.

First, let us present the reduction of $W(E_8)$-invariant
Jacobi forms to the triality invariant Jacobi forms.
We set
\begin{align}
\vecm=\vecm_\tri,\qquad
c=-q^{1/2}\eta^{12},
\end{align}
where $\vecm_\tri$ is given in \eqref{eq:trconf1}.
Then $c^jA_j(\tau,\vecm_\tri),c^jB_j(\tau,\vecm_\tri)$
are expressed in terms of
$\phi_k=\phi_k(\tau,m_1,\ldots,m_4)$,
$\chi_k=\chi_k(\tau,m_1,\ldots,m_4)$
as
\begin{align}
cA_1&=\frac{\Delta(\chi_2+\phi_2)}{6},\nn\\
c^2A_2&=
 \frac{\Delta}{373248}
 \bigl(E_4(5\chi_0^2+48\chi_0\phi_0+18\phi_0^2)\nn\\[.5ex]
&\hspace{2em}
  -E_6(10\chi_0\chi_2+16\chi_0\phi_2+48\chi_2\phi_0+12\phi_0\phi_2)
  +E_4^2(5\chi_2^2+16\chi_2\phi_2+2\phi_2^2)\bigr),\nn\\[.5ex]
c^2B_2&=
 \frac{\Delta}{311040}\bigl(
  -E_6(4\chi_0^2+24\chi_0\phi_0+9\phi_0^2)
  +E_4^2(8\chi_0\chi_2+8\chi_0\phi_2+24\chi_2\phi_0+6\phi_0\phi_2)\nn\\
&\hspace{2em}
  -E_4E_6(4\chi_2^2+8\chi_2\phi_2+\phi_2^2)\bigr)
 -\frac{6\Delta^2\phi_6}{5},\nn\\
c^3A_3&=
 \frac{\Delta E_6}{1003290624}
 \bigl(-E_4(2\chi_0^3+15\chi_0^2\phi_0)
 +E_6(6\chi_0^2\chi_2+5\chi_0^2\phi_2+30\chi_0\chi_2\phi_0)\nn\\[.5ex]
&\hspace{2em}
 -E_4^2(6\chi_0\chi_2^2+10\chi_0\chi_2\phi_2+15\chi_2^2\phi_0)
 +E_4E_6(2\chi_2^3+5\chi_2^2\phi_2)\bigr)\nn\\
&\hspace{1em}
 +\frac{\Delta^2}{580608}
 \bigl(2\chi_0^2\chi_2+3\chi_0^2\phi_2+12\chi_0\chi_2\phi_0
  -24\chi_0\phi_0\phi_2+18\chi_2\phi_0^2-18\phi_0^2\phi_2\nn\\[.5ex]
&\hspace{2em}
  +E_4(2\chi_2^3+3\chi_2^2\phi_2+6\chi_2\phi_2^2+2\phi_2^3)
  -288[E_4(\chi_0+3\phi_0)-E_6(\chi_2+\phi_2)]\phi_6\bigr),\nn\\[.5ex]
c^3B_3&=
 \frac{\Delta E_6}{2866544640}
 \bigl(E_6(2\chi_0^3+15\chi_0^2\phi_0)
  -E_4^2(6\chi_0^2\chi_2+5\chi_0^2\phi_2+30\chi_0\chi_2\phi_0)
\nn\\[.5ex]
&\hspace{2em}
  +E_4E_6(6\chi_0\chi_2^2+10\chi_0\chi_2\phi_2+15\chi_2^2\phi_0)
  -E_6^2(2\chi_2^3+5\chi_2^2\phi_2)\bigr)\nn\\
&\hspace{1em}
 +\frac{\Delta^2}{1658880}
 \bigl(-\chi_0^3-15\chi_0^2\phi_0-36\chi_0\phi_0^2
  -E_6(8\chi_2^3+21\chi_2^2\phi_2+20\chi_2\phi_2^2+4\phi_2^3)
\nn\\[.5ex]
&\hspace{2em}
  +E_4(11\chi_0\chi_2^2+20\chi_0\chi_2\phi_2+12\chi_0\phi_2^2
    +33\chi_2^2\phi_0+36\chi_2\phi_0\phi_2+12\phi_0\phi_2^2)\nn\\[.5ex]
&\hspace{2em}
  +288[E_6(\chi_0+3\phi_0)-E_4^2(\chi_2+\phi_2)]\phi_6\bigr).
\label{eq:ABintri}
\end{align}
Substituting these data into the $E_8$ curve,
one obtains the triality invariant curve.

Due to the equivalence of \eqref{eq:trconf1} and
\eqref{eq:trconf2}, we know that the triality invariant curve
can be viewed as the $F_4$ curve with
slightly altered Wilson line parameters.
This means that the triality invariant curve also takes
the factorized form \eqref{eq:F4canz}.
Therefore, as in the case of the $F_4$ curve,
the data \eqref{eq:ABintri} of $A_j,B_j\ (j\le 3)$
suffice to determine
the triality invariant curve.
By the change of variables
\begin{align}
u\to c^{-1}\left[u-\frac{\chi_0+3\phi_0}{72}
  -\frac{E_6(\chi_2+\phi_2)}{72E_4}\right],\qquad
x\to c^{-2}x,\qquad
y\to c^{-3}y,
\label{eq:uxyE8totr}
\end{align}
we finally obtain
\begin{align}
y^2=4x^3-(f_0u^2+f_1u+f_2)(u-u_*)^2x-(g_0u^3+g_1u^2+g_2u+g_3)(u-u_*)^3,
\label{eq:trcurve1}
\end{align}
where
\begin{align}
\begin{aligned}
f_0&=\frac{E_4}{12},\qquad
f_1 =-\frac{E_6(\chi_2+\phi_2)}{216},\\
f_2&=
 -\frac{E_4\chi_0(\chi_0+12\phi_0)}{20736}
 +\frac{E_6(\chi_0\chi_2+2\chi_0\phi_2+6\chi_2\phi_0)}{10368}
 +\frac{E_4^2(\chi_2-2\phi_2)^2}{62208},\\
g_0&=\frac{E_6}{216},\qquad
g_1 =-\frac{E_4^2(\chi_2+\phi_2)}{2592},\\
g_2&=
 -\frac{E_6\chi_0(\chi_0+6\phi_0)}{124416}
 +\frac{E_4^2(\chi_0\chi_2+\chi_0\phi_2+3\chi_2\phi_0)}{62208}
 +\frac{E_4E_6(\chi_2^2+2\chi_2\phi_2+4\phi_2^2)}{373248}\\
&\hspace{1em}
 -\Delta\phi_6,\\
g_3&=
  \frac{E_6\chi_0^2(\chi_0+12\phi_0)}{8957952}
 -\frac{E_4^2\chi_0(\chi_0\chi_2+2\chi_0\phi_2+12\chi_2\phi_0
        -12\phi_0\phi_2)}{8957952}\\
&\hspace{1em}
 -\frac{E_4E_6(\chi_0\chi_2^2+4\chi_0\phi_2^2+12\chi_2\phi_0\phi_2)}
       {8957952}
 -\frac{(E_4^3-2E_6^2)(\chi_2-2\phi_2)^3}{80621568}\\
&\hspace{1em}
 +\frac{\Delta(\chi_0+3\phi_0)\phi_6}{36},\\
u_*&=\frac{\chi_0+3\phi_0}{36}.
\end{aligned}
\label{eq:trcurve2}
\end{align}

As in the $F_4$ case,
by comparing the above curve with the original Seiberg--Witten curve
\eqref{eq:E8curve},
one immediately obtains the expressions of $a_i,b_j$
at the value \eqref{eq:trconf1} in terms of $\phi_k,\chi_l$.
More specifically, using the change of variables \eqref{eq:uxyE8totr}
inversely as
\begin{align}
u\to cu+\frac{\chi_0+3\phi_0}{72}+\frac{E_6(\chi_2+\phi_2)}{72E_4},
\qquad
x\to c^2x,\qquad
y\to c^3 y,
\end{align}
one can transform the curve \eqref{eq:trcurve1}--\eqref{eq:trcurve2}
into the canonical form \eqref{eq:E8curve}.
Note again that the expressions of $A_i,B_j$
in terms of $a_k,b_l$ are explicitly given in \cite{Sakai:2022taq}.

\subsection{5d limit}

Let us now consider the limit of $q\to 0$,
namely the limit of $T^2$ shrinking to $S^1$.
In this limit, the triality invariant Jacobi forms become
\begin{align}
\begin{aligned}
 \phi_0&=\frac{2}{3}(w_1+w_2+w_4)+32,\qquad
 \phi_2=2(w_1+w_2+w_4)-48,\\
 \phi_6&=4w_3
-\frac{2}{9}(w_1^2+w_2^2+w_4^2)
-\frac{1}{9}(w_1w_2+w_2w_4+w_4w_1)
-\frac{8}{3}(w_1+w_2+w_4)
+32,\\
 \chi_0&=\chi_2=4w_4-2(w_1+w_2),
\end{aligned}
\end{align}
where $w_i$ are the Weyl orbit characters of $D_4$
defined in \eqref{eq:WOCD4}.
Substituting these into \eqref{eq:trcurve1}--\eqref{eq:trcurve2},
the curve becomes
\begin{align}
y^2=4x^3-fx-g
\label{eq:fgcurve}
\end{align}
with
\begin{align}
\begin{aligned}
f&=\frac{1}{15552}(6u-w_4+8)^2(6u-w_4-16)^2,\\
g&=\frac{1}{10077696}(6u-w_4+8)^3(6u-w_4-16)^3.
\end{aligned}
\end{align}
One sees that the discriminant of the curve identically vanishes 
\begin{align}
D=f^3-27g^2=0.
\end{align}
This means that the Seiberg--Witten curve degenerates
everywhere in the Coulomb branch moduli space parametrized by $u$.

The physical interpretation is as follows.
Recall that the Seiberg--Witten curve of the form \eqref{eq:fgcurve}
describes the gauge coupling
$\tilde{\tau}=\frac{\theta}{\pi}+\frac{8\pi i}{g_\mathrm{eff}^2}$
of the effective 4d $\mathcal{N}=2\ \grp{U}(1)$ gauge theory
through the relation
\begin{align}
j(\tilde{\tau})=\frac{1728f^3}{D},
\label{eq:jinv}
\end{align}
where $j=E_4^3/\Delta$.
In the moduli space except at the two common roots of $f$ and $g$,
one has $D=0$ and $f\ne 0$,
which corresponds to $\tilde{\tau}=i\infty$.
At the two exceptional points, $j(\tilde{\tau})$ is undetermined,
but $\tilde\tau(u)$ should be continuous and therefore one has
$\tilde{\tau}=i\infty$ everywhere in the moduli space
including at these points.
This means that $g_\mathrm{eff}=0$ identically,
i.e.~the effective 4d theory is a free $\grp{U}(1)$ theory.
In this sense the low-energy theory is trivial.

Note, however, that if one introduces the Omega background
\cite{Nekrasov:2002qd},
the supersymmetric index exhibits a non-trivial structure.
See \cite[Section~4.1]{Sakai:2014hsa}
for the discussion in the case of $m_i=0$.
It would be interesting to see how the index is deformed
for general $m_i$.

The above limit does not correspond to
any of conventional 5d rank-one gauge theories
on $\bbR^4\times S^1$.
This can be seen by inspecting
the singularities in the moduli space.
The forms of $f$ and $g$ imply that there are
two $D_4$ singularities at the common roots of $f$ and $g$,
whereas there is no singularity at $u=\infty$.
Among 5d $\mathcal{N}=1$ rank-one gauge theories \cite{Seiberg:1996bd},
there are only two theories that exhibit no singularity at $u=\infty$:
One is the pure $\grp{U}(1)$ theory,
but this is not identical to our 5d limit,
because this theory has no singularities in the moduli space.
The other is the $\grp{SU}(2)$ theory with
eight fundamental matters.
This does not correspond to our 5d limit either,
because in the massless limit
the theory has an $\alg{so}(16)$ global symmetry,
which induces a $D_8$ singularity in the moduli space,
but our 5d theory only has two isolated $D_4$ singularities
even in the massless limit.

From a field theory point of view,
it is interesting that our triality invariant model
becomes almost trivial in the 5d limit
but nevertheless has a non-trivial 4d limit,
as we will see below.

\subsection{4d limit}

Recall that the Seiberg--Witten curve
for the $N_\mathrm{f}=4$ theory is given by \cite{Seiberg:1994aj}
\begin{align}
y^2=
 4[W_1W_2W_3+A(W_1S_1(e_2-e_3)+W_2S_2(e_3-e_1)+W_3S_3(e_1-e_2))-A^2N],
\label{eq:Nf4curve1}
\end{align} 
where $e_j$ are given in \eqref{eq:ehdef} and
\begin{align}
\begin{aligned}
A&=(e_1-e_2)(e_2-e_3)(e_3-e_1),\\
W_k&=x-e_ku-\frac{e_k^2}{2}\sum_i m_i^2,\\
S_1&=\frac{1}{12}\sum_{i>j}m_i^2m_j^2-\frac{1}{24}\sum_i m_i^4,\\
S_2&=-\frac{1}{2}\prod_i m_i-\frac{1}{24}\sum_{i>j}m_i^2m_j^2
     +\frac{1}{48}\sum_i m_i^4,\\
S_3&=\frac{1}{2}\prod_i m_i-\frac{1}{24}\sum_{i>j}m_i^2m_j^2
     +\frac{1}{48}\sum_i m_i^4,\\
N&=\frac{3}{16}\sum_{i>j>k}m_i^2m_j^2m_k^2
   -\frac{1}{96}\sum_{i\ne j}m_i^2m_j^4
   +\frac{1}{96}\sum_i m_i^6.
\end{aligned}
\label{eq:Nf4curve2}
\end{align}
The Weierstrass form of this curve is obtained immediately by
performing a translation of $x$
in such a way that the quadratic term in $x$ vanishes.
More specifically, this is achieved
by replacing $x$ with $x+(E_4/144)\sum_j m_j^2$.

Let us now show that our 6d triality invariant
curve \eqref{eq:trcurve1}--\eqref{eq:trcurve2}
reduces to this Weierstrass form.
The 4d limit corresponds to the small mass limit.
Therefore we expand quantities in the 6d curve in $m_j$.
The 6d curve depends on $m_j$ through
$T_k=\prod_{j=1}^4\varth_k(m_j,\tau)$ and $\sum_{j=1}^4\wp(m_j)$.
By using the formulas in Appendix~\ref{app:functions},
we see that they are expanded as
\begin{align}
\begin{aligned}
\sum_{j=1}^4\wp(m_j)
 &=4\pi^2\left[\sum_j \tm_j^{-2}\right.\\
 &\hspace{3.5em}\left.
  +\frac{E_4}{240}\sum_j \tm_j^2+\frac{E_6}{6048}\sum_j \tm_j^4
  +\frac{E_4^2}{172800}\sum_j \tm_j^6 +O(\tm_j^8)\right],\\
T_1(\vecm)
 &=\eta^{12}\left[\prod_j\tm_j\right]\\
 &\hspace{1em}\times
 \exp\left[
  -\frac{E_2}{24}\sum_j \tm_j^2
  -\frac{E_4}{2880}\sum_j \tm_j^4
  -\frac{E_6}{181440}\sum_j \tm_j^6+O(\tm_j^8)
 \right],\\
T_{k+1}(\vecm)
 &=\varth_{k+1}^4
 \exp\left[
  \left(-\frac{E_2}{24}-\frac{e_k}{2}\right)\sum_j \tm_j^2
  +\left(\frac{E_4}{576}-\frac{e_k^2}{4}\right)\sum_j \tm_j^4\right.\\
 &\hspace{6em}\left.
  +\left(\frac{E_6}{12960}+\frac{E_4e_k}{480}-\frac{e_k^3}{6}
   \right)\sum_j \tm_j^6
  +O(\tm_j^8)
     \right],
\end{aligned}
\end{align}
where $\tm_j=2\pi m_j$.
Using these expansions together with
the expressions \eqref{eq:D4gen}
of the fundamental $W(D_4)$-invariant Jacobi forms,
we obtain the expansion of the 6d curve in $m_j$.
Let us then rescale the variables as
\begin{align}
m_j\to \frac{L}{2\pi} m_j,\qquad
x\to -16L^2x,\qquad
y\to 64i L^3y,\qquad
u\to 4L^2u
\end{align}
and expand the equation of
the curve \eqref{eq:trcurve1} in $L$.
We see that all the terms of the order of $L^n\ (n<6)$
cancel out.
The first non-zero part appears at the order of $L^6$, 
and this is precisely the Weierstrass form of the $N_\mathrm{f}=4$
curve \eqref{eq:Nf4curve1}!

\section{Conclusions and outlook}\label{sec:conclusion}

In this paper we have studied two concrete examples of
the E-string theory on $\bbR^4\times T^2$ with Wilson lines,
one exhibits a $W(F_4)$ symmetry and the other exhibits
a $D_4$ triality symmetry.
We have explicitly shown the emergence of these symmetries
by means of the automorphisms
\eqref{eq:WeylE8inv}--\eqref{eq:SL2Zinv}
known for the case of the most general Wilson lines.

In the first example, we have constructed
four different explicit forms of the $F_4$ curve, namely
\eqref{eq:F4curve},
\eqref{eq:F4mucurve}--\eqref{eq:F4mucoeff},
\eqref{eq:thermocurve}
and
\eqref{eq:nonellF4}.
We have shown that they are all equivalent.
The first two expressions are written in terms of
$W(F_4)$-invariant Jacobi forms and thus manifestly symmetric.
The last two expressions are not manifestly symmetric,
but are very concise as one can see.
We have derived these last two expressions from 
the Nekrasov-type formula proposed in \cite{Sakai:2012ik}.
The derivation, together with the equivalence mentioned above,
serves as a proof of the Nekrasov-type formula
in the presence of four general mass parameters.
This generalizes the earlier proof \cite{Ishii:2013nba}
performed in certain special cases.

The second example is the $D_4$ triality invariant model
specified by the twisted Wilson line configuration
\eqref{eq:trconf1}, \eqref{eq:trconf2} or \eqref{eq:trconf3}.
Concerning this model,
we have clarified the discrete automorphism of the
4d $N_\mathrm{f}=4$ theory
and introduced the notion of triality invariant Jacobi forms.
We have then expressed the Seiberg--Witten curve
\eqref{eq:trcurve1}--\eqref{eq:trcurve2}
in terms of them.
We have shown explicitly that this curve reduces
precisely to that of the 4d $N_\mathrm{f}=4$ theory
in the limit of $T^2$ shrinking to zero size.

There are several directions for further studies.
It is definitely interesting to study the supersymmetric
indices in these two examples.
For the E-string theory with general Wilson line parameters $\vecm$,
there are several systematic studies
of the topological string partition functions
\cite{Sakai:2011xg,Kim:2015jba,Kim:2017jqn,Huang:2013yta,
DelZotto:2017mee}
and the elliptic genera \cite{Kim:2014dza,Cai:2014vka}.
In particular, in
\cite{Sakai:2011xg,Huang:2013yta,DelZotto:2017mee,Kim:2014dza,
Cai:2014vka}
the results are already expressed in terms of the $W(E_8)$-invariant
Jacobi forms $A_i,B_j$ given in \eqref{E8AB}.
As explained in the main text, given the Seiberg--Witten curves
\eqref{eq:F4curve} and \eqref{eq:trcurve1}
one immediately obtains the data of the forms
\eqref{eq:ABinF4} and \eqref{eq:ABintri} for all $A_i,B_j$.
The supersymmetric indices are then obtained
by simply substituting these data into the above known results
expressed in terms of $A_i,B_j$.
In particular,
it is interesting to see how the topological string partition
function of the $D_4$ triality invariant model
is related to the Nekrasov partition function
of the $N_\mathrm{f}=4$ theory.
This should be given as a suitable generalization of
the correspondence studied previously in the massless case
\cite{Sakai:2014hsa}.
From a broader perspective, it would also be interesting to
consider the $D_4$ triality invariant deformation of
the elliptic genera of the $E_8\times E_8$ heterotic string
theory \cite{Haghighat:2014pva,Kim:2023glm}. It would be worth
understanding how our results are related to recent studies
such as the twisted elliptic genera of \cite{Lee:2022uiq}
and the bimodular forms of \cite{Aspman:2021evt}.

Another intriguing subject is the quantum Seiberg--Witten curve.
In this paper we have treated the Seiberg--Witten curve
mainly in a manifestly symmetric form.
However, when the symmetry is large,
a curve of this form inevitably has intricate dependence
on the Wilson line parameters
and is not suitable for the quantization.
Quantum Seiberg--Witten curves for the E-string theory
have been studied in
\cite{Moriyama:2020lyk,Moriyama:2021mux,Chen:2021ivd},
where different forms of the classical curve,
more akin to \eqref{eq:nonellF4}, have been employed.
Regarding this, it is worth noting that
the $D_4$ triality invariant curve
\eqref{eq:trcurve1}--\eqref{eq:trcurve2} can also be expressed
in the following simple form:
\begin{align}
u(w^2-1)\eta^6\varth_1(2z,\tau)^2
 =4\prod_{j=1}^4[\varth_j(z+\mu_j,\tau)\varth_j(z-\mu_j,\tau)].
\end{align}
This is obtained by simply substituting \eqref{eq:trconf3}
into the $F_4$ curve \eqref{eq:nonellF4}
and then taking account of the renormalization of Jacobi forms
due to the translation \eqref{eq:trconf1-2rel}
as well as the rescaling \eqref{eq:uxyE8totr}.
This form is neither symmetric nor elliptic,
but is probably more suitable for quantizing the curve
along the lines of \cite{Chen:2021ivd}.

In this paper we have introduced the notion of
$D_4$ triality invariant Jacobi forms.
As far as we know, Jacobi forms of this sort,
i.e.~those for which the underlying modular group is
non-trivially combined with the other automorphism group
have never been considered in the literature before.
As mentioned in section~\ref{sec:trJacobi},
the ring of triality invariant Jacobi forms is not
a polynomial algebra, as in the case of
$W(E_8)$-invariant Jacobi forms
\cite{Wang:2018fil,Sun:2021ije,Sakai:2022taq}.
The structure of the ring deserves to be investigated further.
We will report some results on this problem in a forthcoming paper
\cite{Sakai:2023}.

\vspace{2ex}

\begin{center}
  {\bf Acknowledgments}
\end{center}

This work was supported in part by JSPS KAKENHI Grant Number 19K03856.



\appendix
\renewcommand{\theequation}{\Alph{section}.\arabic{equation}}

\section{Special functions}\label{app:functions}

The Jacobi theta functions are defined as
\begin{align}
\begin{aligned}
\varth_1(z,\tau)&:=
 i\sum_{n\in\bbZ}(-1)^n y^{n-1/2}q^{(n-1/2)^2/2},\\
\varth_2(z,\tau)&:=
  \sum_{n\in\bbZ}y^{n-1/2}q^{(n-1/2)^2/2},\\
\varth_3(z,\tau)&:=
  \sum_{n\in\bbZ}y^n q^{n^2/2},\\
\varth_4(z,\tau)&:=
  \sum_{n\in\bbZ}(-1)^n y^n q^{n^2/2},
\end{aligned}
\end{align}
where
\begin{align}
y=e^{2\pi i z},\qquad q=e^{2\pi i\tau}
\end{align}
and $z\in\bbC, \tau\in\bbH$.
We often abbreviate $\varth_k(0,\tau)$
as $\varth_k(\tau)$ or $\varth_k$.

The Dedekind eta function is defined as
\begin{align}
\eta(\tau):=q^{1/24}\prod_{n=1}^\infty(1-q^n).
\end{align}
The Eisenstein series are given by
\begin{align}
E_{2n}(\tau)
 =1-\frac{4n}{B_{2n}}\sum_{k=1}^{\infty}\frac{k^{2n-1}q^k}{1-q^k}
\end{align}
for $n\in\bbZ_{>0}$.
The Bernoulli numbers $B_k$ are defined by
\begin{align}
\frac{x}{e^x-1}=\sum_{k=0}^\infty\frac{B_k}{k!}x^k.
\end{align}
We often abbreviate $\eta(\tau),\,E_{2n}(\tau)$ as $\eta,\,E_{2n}$
respectively.

The Weierstrass elliptic function is defined as
\begin{align}
\begin{aligned}
\wp(z;2\omega_1,2\omega_3)
  &:= \frac{1}{z^2}
  +\sum_{(m,n)\in\bbZ^2_{\ne (0,0)}}
  \left[\frac{1}{(z-\Omega_{m,n})^2}
    -\frac{1}{{\Omega_{m,n}}^2}\right],
\end{aligned}
\end{align}
where $\Omega_{m,n}=2m\omega_1 + 2n\omega_3$.
In this paper we set
\begin{align}
2\omega_1=1,\qquad 2\omega_3=\tau
\label{eq:omegavalues}
\end{align}
and use the abbreviation
\begin{align}
\wp(z)=\wp(z;1,\tau).
\end{align}

One can show (see e.g.~\cite[Appendix B]{Sakai:2012zq}) that
the Weierstrass elliptic function and
the Jacobi theta functions are expanded as
\begin{align}
\begin{aligned}
\wp(z)&=
  4\pi^2\left(\frac{1}{\tz^2}+\sum_{n=1}^\infty c_{n}\tz^{2n}\right),
\qquad
\tz=2\pi z,\\
\varth_1(z,\tau)&=
 \eta^3\tz\exp\left(
  -\frac{1}{24}E_2 \tz^2
  -\sum_{n=1}^\infty\frac{c_n}{(2n+1)(2n+2)}\tz^{2n+2}\right),\\
\varth_{k+1}(z,\tau)&=
 \varth_{k+1}\exp\left(
  -\frac{1}{24}E_2 \tz^2
  -\sum_{n=1}^\infty\frac{d_n(e_k)}{(2n)!}\tz^{2n}\right)
\qquad (k=1,2,3),
\end{aligned}
\end{align}
where $e_k\ (k=1,2,3)$ are defined in \eqref{eq:ehdef}
and the coefficients $c_n,d_n(x)$ are determined
by the recurrence relations
\begin{align}
\begin{aligned}
c_1 &= \frac{E_4}{240},\qquad
c_2 =       \frac{E_6}{6048},\\
c_n &= \frac{3}{(n-2)(2n+3)}\sum_{k=1}^{n-2}c_k c_{n-k-1}
\qquad (n\ge 3),\\
d_1(x)&=x,\qquad
d_2(x)= 6x^2-\frac{E_4}{24},\\
d_n(x)
&=d_2(x) \partial_x d_{n-1}(x)
 +\left(4x^3-\frac{E_4}{12}x-\frac{E_6}{216}\right)
 \partial_x^2 d_{n-1}(x)
\qquad (n\ge 3).
\end{aligned}
\label{cnrecur}
\end{align}
%


\renewcommand{\section}{\subsection}
\renewcommand{\refname}


{\bf References}

\begin{thebibliography}{100}

\bibitem{Ganor:1996mu}
O.~J.~Ganor and A.~Hanany,
``Small $E_8$ Instantons and Tensionless Non Critical Strings,''
Nucl. Phys. B \textbf{474} (1996), 122--140
[arXiv:hep-th/9602120 [hep-th]].

\bibitem{Seiberg:1996vs}
N.~Seiberg and E.~Witten,
``Comments on String Dynamics in Six Dimensions,''
Nucl. Phys. B \textbf{471} (1996), 121--134
[arXiv:hep-th/9603003 [hep-th]].

\bibitem{Seiberg:1994rs}
N.~Seiberg and E.~Witten,
``Electric-Magnetic Duality, Monopole Condensation, And Confinement in
$N=2$ Supersymmetric Yang-Mills Theory,''
Nucl. Phys. B \textbf{426} (1994), 19--52
[erratum: Nucl. Phys. B \textbf{430} (1994), 485-486]
[arXiv:hep-th/9407087 [hep-th]].

\bibitem{Seiberg:1994aj}
N.~Seiberg and E.~Witten,
``Monopoles, Duality and Chiral Symmetry Breaking in N=2 Supersymmetric
QCD,''
Nucl. Phys. B \textbf{431} (1994), 484--550
[arXiv:hep-th/9408099 [hep-th]].

\bibitem{Ganor:1996xd}
O.~J.~Ganor,
``Toroidal Compactification of Heterotic 6D Non-Critical Strings Down
to Four Dimensions,''
Nucl. Phys. B \textbf{488} (1997), 223--235
[arXiv:hep-th/9608109 [hep-th]].

\bibitem{Ganor:1996pc}
O.~J.~Ganor, D.~R.~Morrison and N.~Seiberg,
``Branes, Calabi--Yau Spaces, and Toroidal Compactification of
the $N$=1 Six-Dimensional $E_8$ Theory,''
Nucl. Phys. B \textbf{487} (1997), 93--127
[arXiv:hep-th/9610251 [hep-th]].

\bibitem{Minahan:1997ch}
J.~A.~Minahan, D.~Nemeschansky and N.~P.~Warner,
``Investigating the BPS Spectrum of Non-Critical $E_n$ Strings,''
Nucl. Phys. B \textbf{508} (1997), 64--106
[arXiv:hep-th/9705237 [hep-th]].

\bibitem{Eguchi:2002fc}
T.~Eguchi and K.~Sakai,
``Seiberg--Witten Curve for the $E$-String Theory,''
JHEP \textbf{05} (2002), 058
[arXiv:hep-th/0203025 [hep-th]].

\bibitem{Eguchi:2002nx}
T.~Eguchi and K.~Sakai,
``Seiberg--Witten Curve for $E$-String Theory Revisited,''
Adv. Theor. Math. Phys. \textbf{7} (2003) no.3, 419--455
[arXiv:hep-th/0211213 [hep-th]].

\bibitem{Minahan:1996cj}
J.~A.~Minahan and D.~Nemeschansky,
``Superconformal Fixed Points with $E_n$ Global Symmetry,''
Nucl. Phys. B \textbf{489} (1997), 24--46
[arXiv:hep-th/9610076 [hep-th]].

\bibitem{Sakai:2012ik}
K.~Sakai,
``Seiberg--Witten prepotential for E-string theory and global
symmetries,''
JHEP \textbf{09} (2012), 077
[arXiv:1207.5739 [hep-th]].

\bibitem{Haghighat:2018dwe}
B.~Haghighat, J.~Kim, W.~Yan and S.~T.~Yau,
``D-type fiber-base duality,''
JHEP \textbf{09} (2018), 060
[arXiv:1806.10335 [hep-th]].

\bibitem{Chen:2021ivd}
J.~Chen, B.~Haghighat, H.~C.~Kim, M.~Sperling and X.~Wang,
``E-string Quantum Curve,''
Nucl. Phys. B \textbf{973} (2021), 115602
[arXiv:2103.16996 [hep-th]].

\bibitem{vandeBult2009}
F.~J.~van de Bult,
``An elliptic hypergeometric integral with $W(F_4)$ symmetry,''
Ramanujan J. \textbf{25} (2011), 1--20
[arXiv:0909.4793 [math.CA]].

\bibitem{Gadde:2009kb}
A.~Gadde, E.~Pomoni, L.~Rastelli and S.~S.~Razamat,
``S-duality and 2d Topological QFT,''
JHEP \textbf{03} (2010), 032
[arXiv:0910.2225 [hep-th]].

\bibitem{Kim:2017toz}
H.~C.~Kim, S.~S.~Razamat, C.~Vafa and G.~Zafrir,
``E-String Theory on Riemann Surfaces,''
Fortsch. Phys. \textbf{66} (2018) no.1, 1700074
[arXiv:1709.02496 [hep-th]].

\bibitem{Hollowood:2003cv}
T.~J.~Hollowood, A.~Iqbal and C.~Vafa,
``Matrix models, geometric engineering and elliptic genera,''
JHEP \textbf{03} (2008), 069
[arXiv:hep-th/0310272 [hep-th]].

\bibitem{Nekrasov:2002qd}
N.~A.~Nekrasov,
``Seiberg-Witten prepotential from instanton counting,''
Adv. Theor. Math. Phys. \textbf{7} (2003) no.5, 831--864
[arXiv:hep-th/0206161 [hep-th]].

\bibitem{Nekrasov:2003rj}
N.~Nekrasov and A.~Okounkov,
``Seiberg-Witten theory and random partitions,''
Prog. Math. \textbf{244} (2006), 525--596
[arXiv:hep-th/0306238 [hep-th]].

\bibitem{Ishii:2013nba}
T.~Ishii and K.~Sakai,
``Thermodynamic limit of the Nekrasov-type formula for E-string
theory,''
JHEP \textbf{02} (2014), 087
[arXiv:1312.1050 [hep-th]].

\bibitem{EichlerZagier}
M.~Eichler and D.~Zagier,
``The Theory of Jacobi forms,''
Prog.\ in Math.\ \textbf{55}, Birkh\"auser-Verlag, 1985.

\bibitem{Wirthmuller}
K.~Wirthm\"uller,
``Root systems and Jacobi forms,''
Comp.\ Math.\  \textbf{82} (1992) 293--354.

\bibitem{Satake:1993cp}
I.~Satake,
``Flat structure for the simple elliptic singularity of type
$\widetilde{\mathrm{E_6}}$ and Jacobi form,''
Proc. Japan Acad. Ser. A Math. Sci. \textbf{69} (1993) no.7, 247--251
[arXiv:hep-th/9307009 [hep-th]].

\bibitem{BertolaThesis}
M.~Bertola,
``Jacobi groups, Jacobi forms and their applications,''
PhD Thesis, SISSA, Trieste, 1999.

\bibitem{Bertola1}
M.~Bertola,
``Frobenius manifold structure on orbit space of Jacobi groups;
Part I,''
Differ.\ Geom.\ Appl.\ \textbf{13} (2000) 19--41.

\bibitem{Sakai:2017ihc}
K.~Sakai,
``$E_n$ Jacobi forms and Seiberg--Witten curves,''
Commun. Num. Theor. Phys. \textbf{13} (2019), 53--80
[arXiv:1706.04619 [hep-th]].

\bibitem{Adler:2019ysr}
D.~Adler and V.~Gritsenko,
``The $D_8$-tower of weak Jacobi forms and applications,''
J. Geom. Phys. \textbf{150} (2020), 103616
[arXiv:1910.05226 [math.AG]].

\bibitem{Adler:2021}
D.~Adler,
``The structure of the algebra of weak Jacobi forms for the root system
$F_4$,''
Funct. Anal. Its Appl. \textbf{54} (2020) 155--168
[arXiv:2007.07116 [math.AG]].

\bibitem{Wang:2018fil}
H.~Wang,
``Weyl invariant $E_8$ Jacobi forms,''
Commun. Num. Theor. Phys. \textbf{15} (2021) no.3, 517--573
[arXiv:1801.08462 [math.NT]].

\bibitem{Sun:2021ije}
K.~Sun and H.~Wang,
``Weyl invariant $E_8$ Jacobi forms and $E$-strings,''
Commun. Num. Theor. Phys. \textbf{17} (2023) no.3, 553--582
[arXiv:2109.10578 [math.NT]].

\bibitem{Sakai:2022taq}
K.~Sakai,
``Algebraic construction of Weyl invariant $E_8$ Jacobi forms,''
Journal of Number Theory \textbf{244} (2023), 42--62
[arXiv:2201.06895 [math.NT]].

\bibitem{DelZotto:2017mee}
M.~Del Zotto, J.~Gu, M.~X.~Huang, A.~K.~Kashani-Poor, A.~Klemm
and G.~Lockhart,
``Topological Strings on Singular Elliptic Calabi-Yau 3-folds
and Minimal 6d SCFTs,''
JHEP \textbf{03} (2018), 156
[arXiv:1712.07017 [hep-th]].

\bibitem{Duan:2020imo}
Z.~Duan, D.~J.~Duque and A.~K.~Kashani-Poor,
``Weyl invariant Jacobi forms along Higgsing trees,''
JHEP \textbf{04} (2021), 224
[arXiv:2012.10427 [hep-th]].

\bibitem{Minahan:1998vr}
J.~A.~Minahan, D.~Nemeschansky, C.~Vafa and N.~P.~Warner,
``$E$-Strings and $N=4$ Topological Yang-Mills Theories,''
Nucl. Phys. B \textbf{527} (1998), 581--623
[arXiv:hep-th/9802168 [hep-th]].

\bibitem{Sakai:2011xg}
K.~Sakai,
``Topological string amplitudes for the local $\frac{1}{2}$K3
surface,''
PTEP \textbf{2017} (2017) no.3, 033B09
[arXiv:1111.3967 [hep-th]].

\bibitem{Ishii:2015iya}
T.~Ishii,
``Comments on the Nekrasov-type formula for E-string theory,''
PTEP \textbf{2015} (2015) no.8, 083B01
[arXiv:1506.05582 [hep-th]].

\bibitem{Sakai:2012zq}
K.~Sakai,
``Seiberg--Witten prepotential for E-string theory and random
partitions,''
JHEP \textbf{06} (2012), 027
[arXiv:1203.2921 [hep-th]].

\bibitem{Seiberg:1996qx}
N.~Seiberg,
``Non-trivial Fixed Points of The Renormalization Group in Six
Dimensions,''
Phys. Lett. B \textbf{390} (1997), 169--171
[arXiv:hep-th/9609161 [hep-th]].

\bibitem{Danielsson:1997kt}
U.~H.~Danielsson, G.~Ferretti, J.~Kalkkinen and P.~Stjernberg,
``Notes on Supersymmetric Gauge Theories in Five and Six Dimensions,''
Phys. Lett. B \textbf{405} (1997), 265--270
[arXiv:hep-th/9703098 [hep-th]].

\bibitem{Mohri:2001zz}
K.~Mohri,
``Exceptional String: Instanton Expansions and Seiberg--Witten Curve,''
Rev. Math. Phys. \textbf{14} (2002), 913--975
[arXiv:hep-th/0110121 [hep-th]].

\bibitem{Sakai:2014hsa}
K.~Sakai,
``A reduced BPS index of E-strings,''
JHEP \textbf{12} (2014), 047
[arXiv:1408.3619 [hep-th]].

\bibitem{Sakai:2023}
K.~Sakai, to appear.

\bibitem{Seiberg:1996bd}
N.~Seiberg,
``Five Dimensional SUSY Field Theories, Non-trivial Fixed Points and
String Dynamics,''
Phys. Lett. B \textbf{388} (1996), 753--760
[arXiv:hep-th/9608111 [hep-th]].

\bibitem{Kim:2015jba}
S.~S.~Kim, M.~Taki and F.~Yagi,
``Tao Probing the End of the World,''
PTEP \textbf{2015} (2015) no.8, 083B02
[arXiv:1504.03672 [hep-th]].

\bibitem{Kim:2017jqn}
S.~S.~Kim and F.~Yagi,
``Topological vertex formalism with O5-plane,''
Phys. Rev. D \textbf{97} (2018) no.2, 026011
[arXiv:1709.01928 [hep-th]].

\bibitem{Huang:2013yta}
M.~X.~Huang, A.~Klemm and M.~Poretschkin,
``Refined stable pair invariants for E-, M- and $[p, q]$-strings,''
JHEP \textbf{11} (2013), 112
[arXiv:1308.0619 [hep-th]].

\bibitem{Kim:2014dza}
J.~Kim, S.~Kim, K.~Lee, J.~Park and C.~Vafa,
``Elliptic Genus of E-strings,''
JHEP \textbf{09} (2017), 098
[arXiv:1411.2324 [hep-th]].

\bibitem{Cai:2014vka}
W.~Cai, M.~x.~Huang and K.~Sun,
``On the Elliptic Genus of Three E-strings and Heterotic Strings,''
JHEP \textbf{01} (2015), 079
[arXiv:1411.2801 [hep-th]].

\bibitem{Haghighat:2014pva}
B.~Haghighat, G.~Lockhart and C.~Vafa,
``Fusing E-strings to heterotic strings: E + E $\rightarrow$ H,''
Phys. Rev. D \textbf{90} (2014) no.12, 126012
[arXiv:1406.0850 [hep-th]].

\bibitem{Kim:2023glm}
H.~C.~Kim, M.~Kim and Y.~Sugimoto,
``Blowup Equations for Little Strings,''
JHEP \textbf{05} (2023), 029
[arXiv:2301.04151 [hep-th]].

\bibitem{Lee:2022uiq}
K.~Lee, K.~Sun and X.~Wang,
``Twisted Elliptic Genera,''
JHEP \textbf{04} (2024), 035
[arXiv:2212.07341 [hep-th]].

\bibitem{Aspman:2021evt}
J.~Aspman, E.~Furrer and J.~Manschot,
``Four flavors, triality, and bimodular forms,''
Phys. Rev. D \textbf{105} (2022) no.2, 025017
[arXiv:2110.11969 [hep-th]].

\bibitem{Moriyama:2020lyk}
S.~Moriyama,
``Spectral Theories and Topological Strings on del Pezzo Geometries,''
JHEP \textbf{10} (2020), 154
[arXiv:2007.05148 [hep-th]].

\bibitem{Moriyama:2021mux}
S.~Moriyama and Y.~Yamada,
``Quantum Representation of Affine Weyl Groups and Associated
Quantum Curves,''
SIGMA \textbf{17} (2021), 076
[arXiv:2104.06661 [math.QA]].

\end{thebibliography}
\end{document}